\documentclass[preprint,showpacs,aps]{revtex4}
\usepackage{graphicx}
\usepackage{dcolumn}
\usepackage{bm}

\def\by#1#2{{\displaystyle {#1}\over \displaystyle {#2}}}
\def\d{{\rm d}}
\preprint {IMSc/2006/03/09}
\begin{document}
\title{Neutrino oscillation probabilities: Sensitivity to parameters} 
\author{D. Indumathi, M.V.N. Murthy, G. Rajasekaran and Nita Sinha}

\affiliation
{The Institute of Mathematical Sciences, Chennai 600 113, India.\\
}
\date{\today}

\begin{abstract}

We study in detail the sensitivity of neutrino oscillation
probabilities to the fundamental neutrino parameters and their possible
determination through experiments. The first part of the paper is
devoted to the broad theme of isolating regions in the neutrino (and
anti-neutrino) energy and propagation length that are sensitive to the
oscillation parameters. Such a study is relevant to neutrinos both from
the Earth's atmosphere or from a neutrino factory. For completeness we
discuss the sensitivity, however small, to the parameters involved in a
three-generation framework, and to the Earth matter density profile. We
then study processes relevant to atmospheric neutrinos which are
sensitive to and allow precision measurements of the mixing angle
$\theta_{23}$ and mass-squared difference $\delta_{32}$ apart from the
mixing angle $\theta_{13}$. Crucial to this analysis is charge
identification; detectors having this capability can isolate these
matter effects. In particular, we address the issue of using matter
effects to determine whether the mixing angle $\theta_{23}$ is maximal,
and, if not, to explore how well its octant can be determined. When
realistic detector resolutions are included, we find that deviations of
about 15\% (20\%) from a maximal value of $\sin^2\theta_{23}=1/2$ can
be measured at 95\% (99\%) CL) provided $\theta_{13}$ is non-zero,
$\sin^2 \theta_{13} \ge 0.015$, and the neutrino mass ordering is
normal, with fairly large exposures of 1000 kton-years.

\end{abstract}


\maketitle

\section{Introduction}

Observations of solar, atmospheric, reactor and long-baseline
neutrinos have provided compelling evidence for neutrino oscillations
\cite{solar,atm,reactor, ktok}. In fact, the latest analysis of the
data from the Super-Kamiokande (Super-K) collaboration \cite{ishi}
presents a clear and unambiguous evidence for the oscillation hypothesis
\cite{theory,mixing} by establishing the first {\it oscillation minimum}
and hence for non-zero (and different) neutrino masses and mixing. In
the meanwhile, the Sudbury Neutrino Observatory (SNO) \cite{solar} has
provided incontrovertible evidence for neutrino oscillation from solar
neutrino data.

Clearly neutrino physics is now entering an era of precision
measurements. The focus from now on will be to precisely determine the
oscillation parameters apart from the observation of the oscillation
pattern beyond the first minimum. Given the reach of present and
proposed experiments, the goal is to understand in detail the
sensitivity of the oscillation probabilities to fundamental neutrino
parameters, which are the mass-squared differences and mixing angles
and phases. It is therefore crucial to choose processes or observables
that maximise the sensitivity to the parameter(s) of interest.

It turns out that the density profile of matter that the neutrinos
pass through before detection naturally accentuates the sensitivity
to the mixing angles $\theta_{13}$, $\theta_{23}$ and the mass squared
difference $\delta_{32}$. These {\em matter effects} have been studied
before, with atmospheric neutrinos, in various contexts, in Refs.~
\cite{petcov,fogli,huber,akhmedov1,tabarelli,gandhi,debu,probir,IM}.
In particular, substantial matter dependences occur in the energy range of
interest only if $\sin^2\theta_{13}$ is different from zero; a non-zero
value of this parameter also opens the window to CP violation in the
lepton sector. Since only an upper bound exists on this parameter so far,
the variation in sensitivity to other parameters can be large or small,
depending on the true value of $\theta_{13}$.

This paper is devoted to the detailed study of oscillation probabilities
of energetic (GeV energy) neutrinos and anti-neutrinos propagating
through the Earth in a three flavour framework using accurate numerical
methods. The paper is divided into two parts: The first part is devoted
to the broad theme of sensitivity of the neutrino (and anti-neutrino)
oscillation probabilities to the fundamental neutrino parameters in
the GeV region. The results of this analysis are equally applicable to
atmospheric neutrinos as well as to neutrino beams from future neutrino
factories. While some of this information is already known, we
emphasise a few surprising variations.



Having determined the various sensitivities, we then use the
results of the first part to maximise matter effects in a study of
atmospheric neutrinos (and anti-neutrinos) using detectors having
charge identification capability and study how best these parameters
can be constrained/determined by such an analysis.


The paper is organised as follows. In section 2, we outline the framework
for the calculation of neutrino probabilities. Though
numerical procedures have been developed to calculate any desired
survival or conversion probability, in this paper we focus on the
probabilities relevant for the accurate calculation of muon-neutrino
(and anti-neutrino) events at a detector.  We also list here the current
limits on the oscillation parameters.

In section 3, we consider, in turn, the variation of the muon neutrino
survival and conversion probabilities as the oscillation parameters are
varied within the allowed limits. The compendium of results listed here
applies generically to any combination of neutrino source and detector
where GeV energy neutrinos traverse the Earth. 

In section 4, we present results for event rates calculated with
atmospheric neutrinos. We also identify in passing interesting regions
in energy--nadir angle space for long-baseline neutrinos. The results are
calculated for up/down ratios of event rates binned as a function
of the zenith angle or equivalently distance $L$ to bring out the maximal
sensitivity to parameters, especially $\theta_{13}$, $\theta_{23}$ and
the sign of $\delta_{32}$. The question of hierarchy from the sign of
$\delta_{32}$ has been discussed in detail before \cite{IM}; we focus
here on the sensitivity of event rates and event ratios to the magnitude
of this parameter. An important open question is the deviation from
maximal of the mixing angle $\theta_{23}$, that is, its octant
sensitivity. This was analysed recently by Choubey and Roy \cite{probir};
we present results here which are complementary to their analysis.
The issue of both the mass hierarchy and the octant of $\theta_{23}$ have
relevance to the structure of neutrino mass models.

In section 5 we present a summary of our results and conclusions; we
also highlight the improvement over the earlier results. Some remarks
on the numerical computation are presented in the appendix. We also
discuss the possible points of departure from earlier calculations,
especially those that have been used in neutrino event generators such 
as {\sc nuance} \cite{nuance}.

\section{Oscillation probabilities}

For completeness we review the basic framework used in evolving the
neutrino states and the calculation of oscillation probabilities.
In a three neutrino framework, the neutrino flavour states $|\nu_\alpha
\rangle$, $\alpha = e, \mu, \tau$, are defined as linear superpositions
of the neutrino mass eigenstates $|\nu_i\rangle$, $i=1,2,3$, with
masses $m_i$ :
\begin{equation}
\vert\nu_\alpha \rangle = \sum_i U_{\alpha i} \vert\nu_i\rangle~.
\end{equation}
The $|\nu_{\alpha}\rangle$ are 
the states produced in association with the charged leptons. 
The $3 \times 3$ unitary matrix $U$ may be parametrised 
\cite{mixing} (ignoring Majorana phases, which
may be included by multiplying $U$ by a diagonal phase matrix involving 
two more phases in addition to $\delta$) as:
\begin{eqnarray}
U = \left(
          \begin{array}{ccc}
          c_{12}c_{13} & s_{12}c_{13} & s_{13}e^{-i\delta}  \\
 -c_{23}s_{12} - s_{23}s_{13}c_{12}e^{i\delta} & c_{23}c_{12} -
s_{23}s_{13}s_{12}e^{i\delta}&  s_{23}c_{13}\\
  s_{23}s_{12} - c_{23}s_{13}c_{12}e^{i\delta}& -s_{23}c_{12} -
c_{23}s_{13}s_{12}e^{i\delta} & c_{23}c_{13} \end{array} \right),
\label{eq:mns}
\end{eqnarray}
where $c_{ij}=\cos\theta_{ij}$, $s_{ij}=\sin\theta_{ij}$ and 
$\delta$
denotes the CP violating (Dirac) phase. The 3 $\times$
3 neutrino mass matrix $M_\nu^2$ in the charged-lepton
mass basis is diagonalised by $U$:
\begin{equation} 
U^\dagger M^2_\nu U = {\rm{diag}}(m_1^2,m_2^2,m_3^2). 
\label{dia}
\end{equation} 
The fundamental neutrino parameters therefore are the mixing angles
$\theta_{ij}$, phase $\delta$, and the mass-squared differences
$\delta_{ij} \equiv m_i^2 - m_j^2$.

The time evolution of the mass eigenstates is governed by the 
Schroedinger equation, 
\begin{equation}
|\nu_i(t)\rangle = e^{-iE_i t}|\nu_i(0)\rangle.
\end{equation}
Consequently the time evolution of the flavour states is given by the 
equation,
\begin{equation}
i\frac{\d}{\d t}\left[ \nu_\alpha \right] =\frac{1}{2E} 
U M_\nu^2 U^\dagger
\left[\nu_\alpha\right]~.
\end{equation}
where $[\nu_{\alpha}]$ is the vector of flavour eigenstates, 
$\left[\nu_{\alpha}\right]^T = 
\left[\vert\nu_e\rangle,\vert\nu_{\mu}\rangle,\vert\nu_{\tau}\rangle \right]$.

The evolution equation in the presence of matter is given by 
\begin{eqnarray}
i\frac{\d}{\d t} \left[ \nu_\alpha \right] = \frac{1}{2E}
\left[U M_{\nu}^2 U^{\dagger} 
 +\left(
          \begin{array}{ccc}
          A & 0 & 0  \\
          0 & 0 & 0  \\
          0 & 0 & 0 \end{array} \right)\right]
\left[ \nu_\alpha \right]~,
\label{eq:ham}
\end{eqnarray}
where the matter term $A$ (ignoring the diagonal neutral current
contribution) is given by
\begin{equation}
A = 2 \sqrt{2} G_F n_e E = 7.63 \times 10^{-5}~{\rm eV}^2~\rho({\rm gm/cc})~
E({\rm GeV})~ \hbox{eV}^2.
\label{densm}
\end{equation}
Here $G_F$ and $n_e$ are the Fermi constant and electron number density
in matter and $\rho$ is the matter density. The evolution equation for 
anti-neutrinos has the sign of $A$ and the phase
$\delta$ reversed. 

In general, the 3-flavour probabilities, in particular, $P_{\mu\mu}$ and
$P_{e\mu}$ of interest here, depend on all the oscillation parameters:
$\theta_{12}$, $\theta_{23}$, $\theta_{13}$, $\delta_{21}$,
$\delta_{32}$, and the CP phase $\delta$.
We will focus everywhere on effects due to propagation in (Earth's)
matter. It is useful to compute these probabilities in the constant
density approximation for comparison with the exact numerical results
below. We have \cite{matprob},
\begin{eqnarray}
P^m_{\mu\mu} & \approx & 1
- \sin^4\theta_{23} \sin^2 2\theta_{13}^{m} \sin^2\Delta_{31}^m~,
\nonumber \\
\ & & \qquad - \sin^2 2\theta_{23} \,
\left[\sin^2\theta_{13}^{m} \sin^2\Delta_{21}^m +
\cos^2\theta_{13}^{m}\sin^2\Delta_{32}^m \right]~, \\  
P_{e\mu} & \approx & \sin^2 \theta_{23} \sin^2 2\theta_{13}^{m}\,
\sin^2 \Delta_{31}^m~,\nonumber ~
\label{eq:p3num}
\end{eqnarray} 
where terms involving $\delta_{21}$ have been ignored \cite{IM} and 
the superscript $m$ refers to mixing angles and mass square
differences in matter. The relevant $L/E$-dependent terms in matter 
are given by,
\begin{eqnarray}
\Delta_{21}^m & = & \frac{1.27 \delta_{32} L}{E}\frac{1}{2}\,
\left[\frac{\sin 2\theta_{13}}{\sin 2\theta_{13,m}} - 1 -
\frac{A}{\delta_{32}} \right]~, \nonumber \\
\Delta_{32}^m & = & \frac{1.27 \delta_{32} L}{E}\frac{1}{2}\,
\left[\frac{\sin 2\theta_{13}}{\sin 2\theta_{13,m}} + 1 +
\frac{A}{\delta_{32}} \right]~, \nonumber \\
\Delta_{31}^m & = & \frac{1.27 \delta_{32} L}{E}\,
\left[\frac{\sin 2\theta_{13}}{\sin 2\theta_{13,m}} \right]~. 
\end{eqnarray}
Note that for sufficiently large values of $A/\delta_{32}$, all the
three scales are of the same order of magnitude including 
$\Delta_{21}^m$ which cannot therefore be neglected.

While $\theta_{13}$ is small, its value in matter can be enhanced:
\begin{equation}
\sin 2\theta_{13}^{m}=\frac{\sin2\theta_{13}}{\sqrt{(\cos2\theta_{13}-
A/\delta_{32})^2+(\sin 2\theta_{13})^2}}~. \nonumber
\end{equation}
In particular, $\sin2\theta_{13}^{m} = 1$ at resonance, when
\begin{equation}
\delta_{32} \,\cos2\theta_{13} = A~.
\label{eq:resonance}
\end{equation}

It is convenient to introduce the scale $\Delta m^2$\cite{foglipar} 
instead of
$\delta_{32}$:
$$
\Delta m^2 \equiv \delta_{32} + 
\delta_{21}/2~=~m_3^2-\frac{1}{2}(m_1^2+m_2^2),
$$
so that a normal or inverted hierarchy is simply indicated by a sign
(and not magnitude) change in this parameter.

Data on neutrino oscillations are available from the following
experiments:
\begin{itemize}

\item A combination of solar neutrino experiments and the KamLAND 
reactor\cite{solar,reactor} 
experiment sets limits on the parameters $\delta_{21}$ and
$\theta_{12}$; it also establishes the mass ordering $m_2 > m_1$.

\item A combination of atmospheric neutrino experiments and the K2K
experiment\cite{atm,ktok} constrains the parameters in the (23) sector:
$\delta_{32}$ (or $\Delta m^2$) and $\theta_{23}$. Note that the sign
of this mass-squared difference as well as the deviation of
$\theta_{23}$ from maximality are not yet determined.

\item The reactor neutrino experiment, {\sc chooz}\cite{reactor}, sets
an upper bound on the effective (13) mixing angle using the above
parametrisation.

\item The CP phase $\delta$ is unknown.

\end{itemize}

A combined analysis of these data \cite{foglipar} places limits on the
oscillation parameters as summarised in Table~\ref{tab:bestfit}.

\begin{table}
\begin{tabular}{|l|r|} \hline
Parameter & \multicolumn{1}{c|}{Best-fit value} \\ \hline
$\delta_{21}$ & $7.92 \, \left(1 \pm 0.09\right) \times 10^{-5}$ eV$^2$ \\
$|\Delta m^2|$  & $2.40 \, \left(1^{+0.21}_{-0.26}\right) \times 
10^{-3}$ eV$^2$ \\
              & \\ \hline
$\sin^2 \theta_{12}$; \quad $\left[\theta_{12} \right]$
 & $0.314 \, \left(1^{+0.18}_{-0.15}\right)$; \quad $\left[34.1^\circ \right]$ \\
$\sin^2 \theta_{13}$; \quad $\left[\theta_{13} \right]$
 & $< 0.032$; \quad $\left[10.3^\circ\right]$ \\
$\sin^2 \theta_{23}$; \quad $\left[\theta_{23} \right]$
 & $0.44 \, \left(1^{+0.41}_{-0.22}\right)$; \quad $\left[ 41.6^\circ \right]$ \\ \hline
\end{tabular}
\caption{Table showing currently accepted \cite{foglipar} best-fit values
of oscillation parameters with 2$\sigma$ errors. In the case of the
mixing angle $\theta_{13}$, a 2$\sigma$ upper bound is shown.}
\label{tab:bestfit}
\end{table}

\section{Sensitivity of oscillation probabilities to parameters}

We now discuss in detail the $\nu_{\mu}$, $\bar{\nu}_{\mu}$ survival
and conversion probabilities as these are of interest here. Note that,
in the absence of oscillations, the number of events in the neutrino
sector is always a factor of two or more larger than in the case of
anti-neutrinos due to the larger neutrino cross-section. We will
comment further on its implication in section 4.

The extraction of neutrino parameters to high precision requires
the determination of neutrino oscillation probabilities in matter to a
high degree of accuracy. Analytically, the effect of non-uniform matter
poses a problem, while, in principle, the oscillation probabilities may
be computed numerically to the requisite precision. Many, sometimes
complicated, analytical formulae exist for neutrino propagation in
vacuum and in matter with either constant or variable density~%
\cite{rajaji,akhmedov,vacuum,matter,linear,expo,approx}.

We use a Runge-Kutta solver to calculate the oscillation probabilities
for various energies and nadir angles. Some technical details are given in
Appendix A. All results presented in this and the following sections
have been obtained  using the density profile of the Earth as given
by the Preliminary Reference Earth Model ({\sc prem}) \cite{prem} and
numerically evolving the flavour eigenstates through Earth's matter. In
particular, the approximate expressions for the probabilities as shown
in Eq.~\ref{eq:p3num} are not used.

\begin{figure}[htp] 
\includegraphics[height = 10cm]{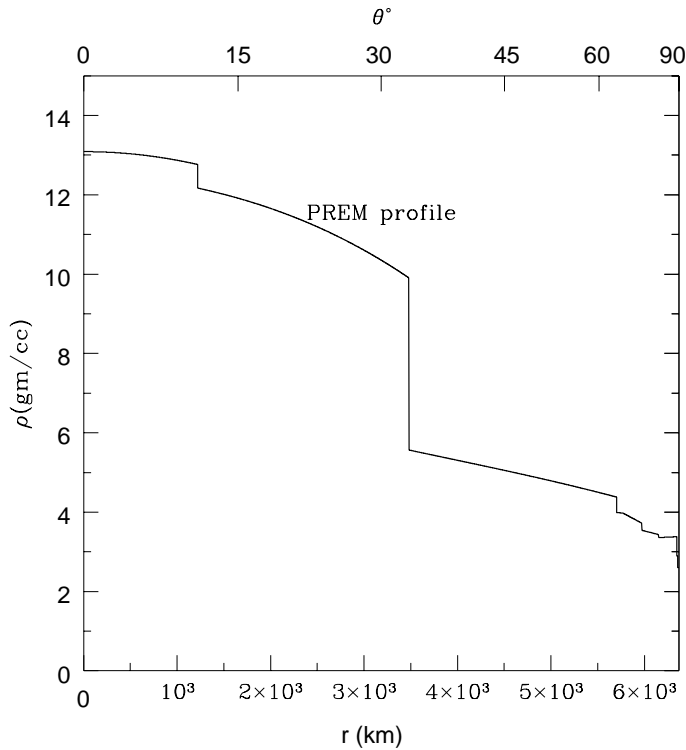}
\caption{Variation of the Earth's density profile (assumed spherically
symmetrical) with distance from the centre (in km) in the PREM
model \cite{prem}. The equivalent largest nadir angle of neutrinos that
just graze a shell of given radius $r$ is also shown.}
\label{fig:prem}
\end{figure}

The matter profile in the {\sc prem} model is shown in
Fig.~\ref{fig:prem} as a function of the radius $r$ from
the centre of the Earth. The density jumps at inner-outer core and
core-mantle transitions are clearly seen. An up-going neutrino with
nadir angle $\theta$ (shown in the upper $x$-axis) traverses all density
zones of radii larger than the corresponding $r$ shown in the lower
$x$-axis. For $\theta=33^\circ$, neutrinos graze the core-mantle
boundary, while for smaller $\theta$, they traverse the core.

We also implicitly assume an iron calorimeter detector with charge
identification capability as in the {\sc minos}\cite{minos} and the
proposed {\sc monolith} \cite{monolith} and {\sc ical/ino} \cite{ino}
detectors. We therefore focus on neutrino energies in the GeV range
which is of relevance to these detectors. We will see that the most
interesting and sensitive region, purely in terms of the probabilities,
is between 2--10 GeV. We look at the sensitivity of the oscillation
probabilities to the fundamental neutrino parameters, namely the mass
squared differences $\delta_{21}$, $\Delta m^2$, the mixing angles
$\theta_{12}$, $\theta_{13}$, $\theta_{23}$ and the CP phase $\delta$.

In order to assess the parameter sensitivity, we consider $2\sigma$
variations from the current best fit value of each parameter while
the others are kept at their best fit values. We shall see that the
probabilities are sensitive mainly to the (23) and (13) mixing angles
and the larger mass-squared difference $\Delta m^2$. When matter
effects are turned on the probabilities are also sensitive to the
sign of this mass-squared difference. There is also a dependence on
the density profile of the Earth. We shall discuss these dependences
one by one. Since matter effects are proportional to $\sin\theta_{13}$,
which is small, we are looking for sub-leading effects. The main role of
matter is to enhance $\sin\theta_{13}$ due to resonance effects. Hence,
when we emphasise a large effect, it is understood to be ``large'' with
respect to this small parameter; matter effects, unfortunately, remain
rather small with respect to the total events/processes we consider and
hence need large exposures for their study.

We shall consider neutrino energies in the range 2--10 GeV, which are
relevant for atmospheric neutrino studies. Typically, we shall hold other
parameters at their known central values when we discuss the variation
with any one parameter. In particular, there is only a marginal/negligible
dependence of these probabilities on $\theta_{12}$ and $\delta m^2_{21}$
and we shall not discuss them further.

In general the normal hierarchy is assumed; this means that all matter
effects are enhanced due to resonance in the neutrino sector, with
smaller effects (and no resonances) in the anti-neutrino sector. Since
we are only discussing probabilities, there is an approximate symmetry
between particle probabilities with normal hierarchy and anti-particle
probabilities with inverted hierarchy \cite{IM}. This symmetry is exact
in the limit that terms containing $\delta_{21}$ can be ignored. Thus we
choose $\Delta m^2$ to be positive (normal hierarchy) while discussing
probabilities. Results with the inverted hierarchy are discussed in the
next section.

\subsection{Variation of probabilities with $\theta_{13}$}

\begin{figure}[htp] 
\includegraphics[width = \textwidth]{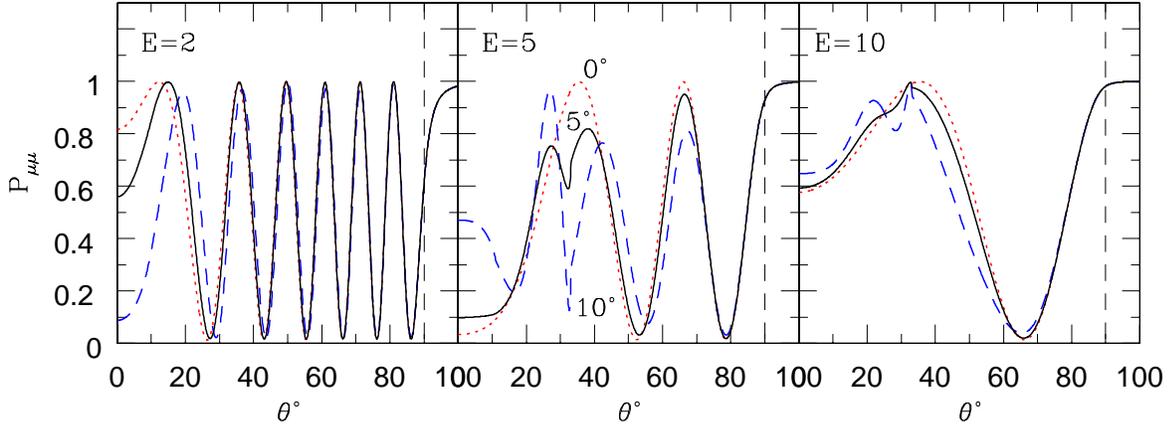}
\caption{Variation of $P_{\mu\mu}$ with $\theta_{13}$ as a function
of nadir angle $\theta$.  Probabilities for $\theta_{13} = 0, 5,
10^\circ$ are shown as dotted, solid and dashed lines respectively.
Sensitivity to $\theta_{13}$ at three different neutrino energies,
$E=2,5,10$ GeV, is shown in the three panels. Other parameters are set
to their best-fit central values (see Table \ref{tab:bestfit}). Normal
hierarchy is assumed.}
\label{fig:13t} 
\end{figure}

\begin{figure}[htp]
\includegraphics[width = \textwidth]{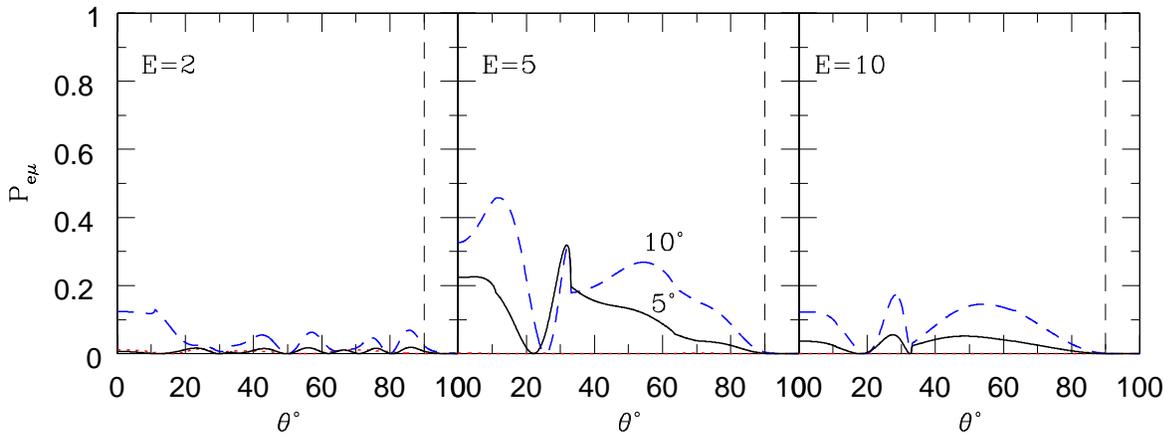}
\caption{Variation of $P_{e\mu}$ with $\theta_{13}$. See caption of
Fig.~\ref{fig:13t}. Although $P_{e\mu}$ is non-zero when $\theta_{13} = 0$,
the non-zero contribution comes from an oscillatory term dependent on
$\delta_{21}$ and is very small/invisible beyond $E \sim 1$ GeV.}
\label{fig:e13t}
\end{figure}

In general, matter effects are significant provided $\theta_{13}$ is
not small. The dependence of matter effects on this parameter has been
exhaustively studied elsewhere \cite{petcov,fogli,
huber,akhmedov1,tabarelli,gandhi,debu,probir,IM} and will not be
discussed in any detail in this paper. We merely show a sample
dependence on this parameter in Figs.~\ref{fig:13t} and ~\ref{fig:e13t}
where $P_{\mu\mu}$ and $P_{e\mu}$ are plotted as a function of the
nadir angle $\theta$ for three energies, $E=2,5,10$ GeV. The variation
in the probabilities as $\theta_{13}$ varies over its 2$\sigma$ allowed
region, $\theta_{13} < 10.3^\circ$, is significant at small nadir
angles for low energies ($E<3$ GeV), at virtually all nadir angles at
larger energies (around $E \sim 5$ GeV), and tapers off at larger
energies (by about $E \sim 10$ GeV) when matter effects become
irrelevant. Notice the effect of core crossing at nadir angle $\theta =
33^\circ$ which is especially visible when $E=5$ GeV when
$\theta_{13}\ne 0$. . 

\begin{figure}[htp]
\includegraphics[width = \textwidth]{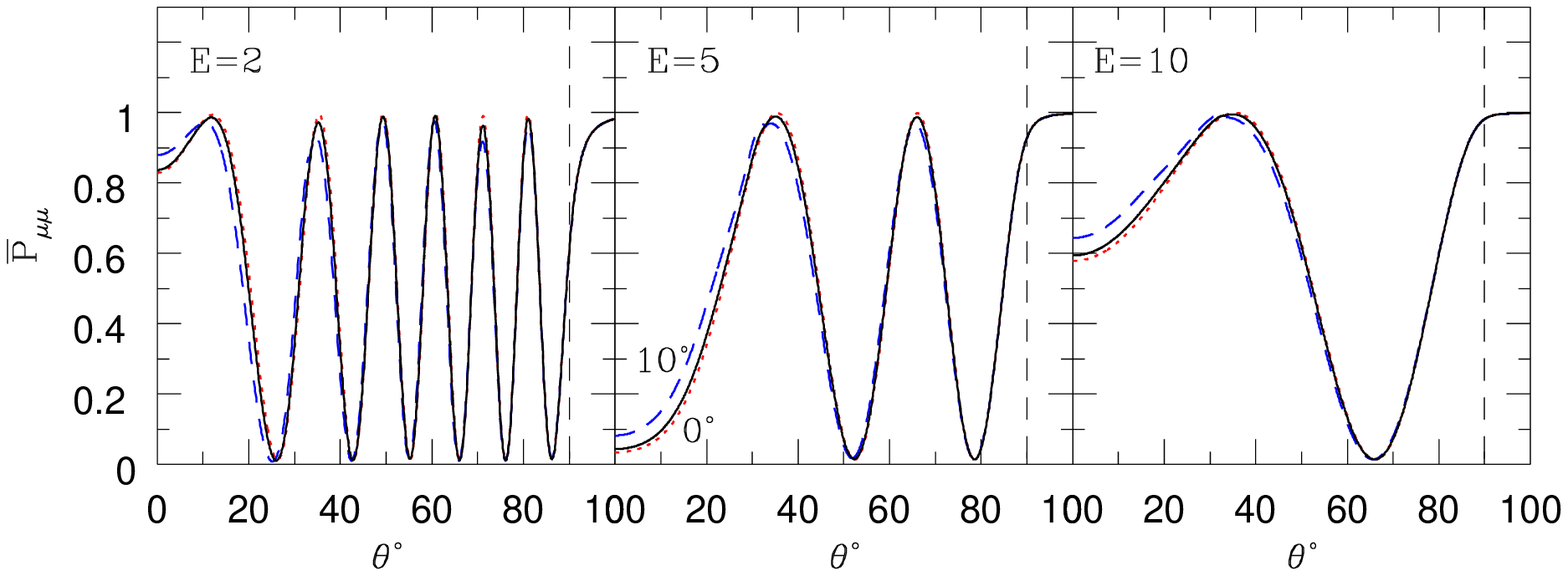}
\caption{Variation of $\overline{P}_{\mu\mu}$ with $\theta_{13}$. See
caption of Fig.~\ref{fig:13t}.}
\label{fig:13bart}
\end{figure}

For completeness we show the relatively weaker sensitivity to this
parameter of the anti-neutrino probability $\overline{P}_{\mu\mu}$ in
Fig.~\ref{fig:13bart}. When the hierarchy is inverted the behaviour of
$P$ and $\overline{P}$ are interchanged.

\subsection{Variation of probabilities with $\Delta m^2$}

\begin{figure}[htp]
\includegraphics[width = \textwidth]{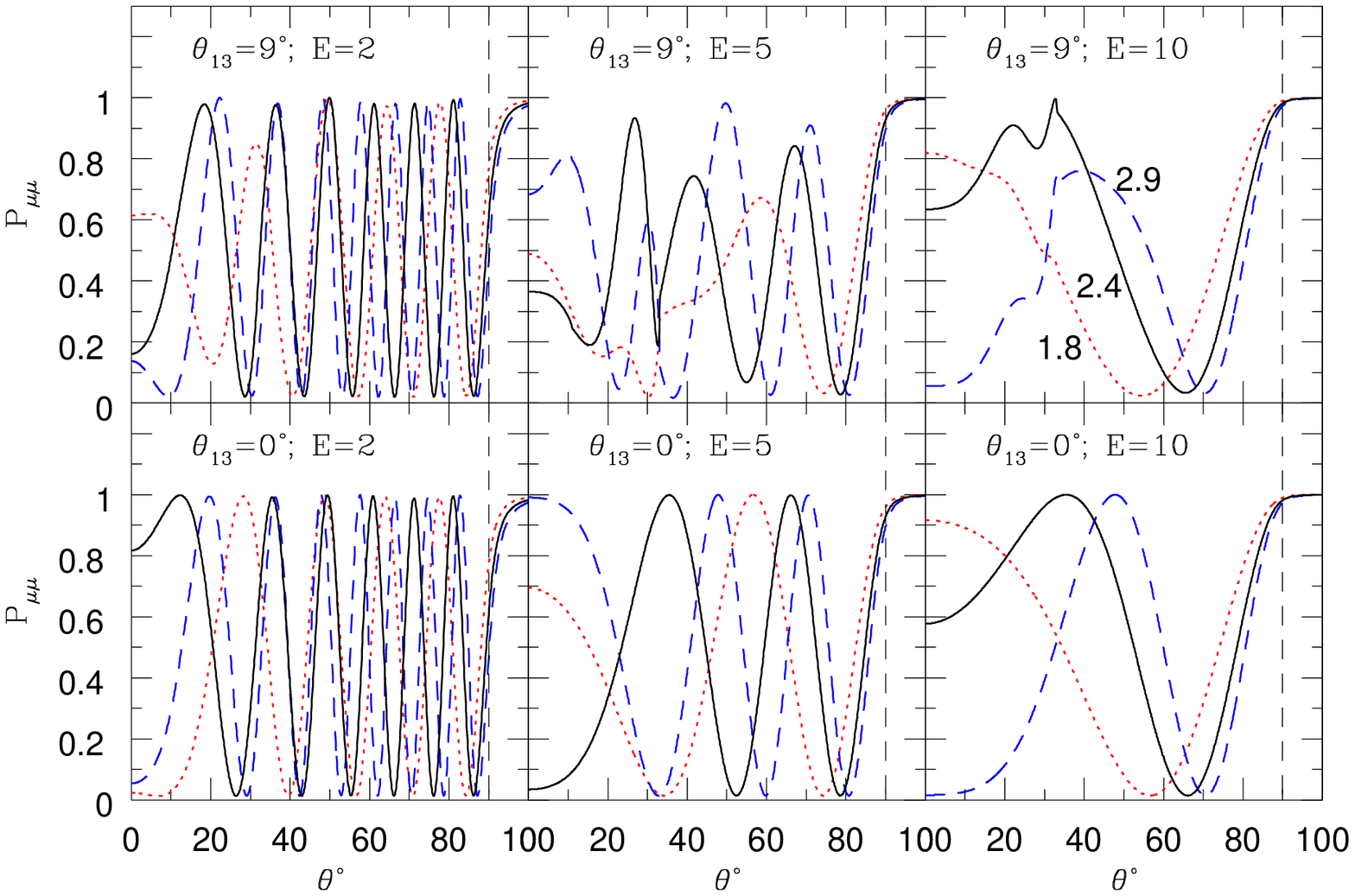}
\caption{Variation of $P_{\mu\mu}$ with $\Delta m^2$ at $\theta_{13} =
9^\circ$ (upper panels) and $\theta_{13} = 0$ (lower panels). $\Delta
m^2$ values used are the 2$\sigma$ allowed variation (dotted line for
$1.8\times 10^{-3}$ and dashed line for $2.9\times 10^{-3}$ eV$^2$) of
this parameter around the solid curve for the best-fit value ($ = 2.4
\times 10^{-3}$ eV$^2$). The values of $\Delta m^2$ in units of
$10^{-3}$ eV$^2$ are marked in the top right panel of the figure.}
\label{fig:32t}
\end{figure}

We begin with $P_{\mu\mu}$, which is extremely sensitive to this
parameter, as can be seen in Fig.~\ref{fig:32t}. The figure shows the
variation of $P_{\mu\mu}$ with the nadir angle $\theta$ at different
neutrino energies, for a 2$\sigma$ variation of $\Delta m^2$: $\Delta
m^2 = (1.8, 2.4, 2.9) \times 10^{-3}$ eV$^2$. The top three panels are
for $\theta_{13}=9^\circ$ and the lower three for $\theta_{13}=0$ at
three different energies. Notice that the smooth oscillations seen
in the lower panels are modified due to matter effects (turned on when
$\theta_{13} \neq 0$) in the upper panel.

In general, the modification is significant in the small nadir angle
bins for lower energies ($E=2$ GeV), and at all nadir angle bins in the
intermediate energy region (represented by the central panel at $E=5$
GeV) when resonance occurs in the mantle. The interplay between the
variation in $\Delta m^2$ (even at the 2$\sigma$ level) and that in
the relatively unknown $\theta_{13}$ is clear from the $E=5$ GeV panels:
the same muon survival probability for upward-going neutrinos ($\theta
=0^\circ$) can come from ($\Delta m^2, \theta_{13}$) both large:
($2.9\times 10^{-3}$ eV$^2, 9^\circ$) or both small: ($1.8\times 10^{-3}$
eV$^2$, $\sim 0^\circ$).

\begin{figure}[htp]
\includegraphics[width = \textwidth]{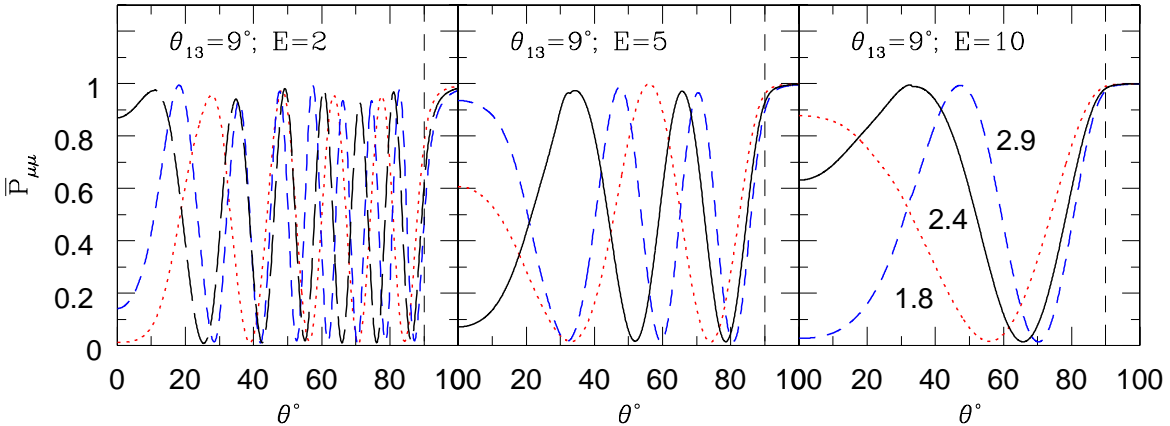}
\caption{Variation of $\overline{P}_{\mu\mu}$ with $\Delta m^2$ at
$\theta_{13} = 9^\circ$. See caption of Fig.~\ref{fig:32t} for details.
The probability is not significantly different at other values of
$\theta_{13}$ and is similar to the neutrino survival probability in
Fig.~\ref{fig:32t} for $\theta_{13}=0$.}
\label{fig:32bart}
\end{figure}

Since matter effects are relatively unimportant for anti-neutrinos (in
the normal hierarchy) the anti-neutrino survival probability at all
allowed $\theta_{13}$ is close to the neutrino survival probability at
$\theta_{13} =0$. This can be seen from Fig.~\ref{fig:32bart} where
$\overline{P}_{\mu\mu}$ is shown. In fact, the difference between the
neutrino and anti-neutrino survival probabilities is a measure of the
matter effect and hence that of $\theta_{13}$ and the sign of $\Delta
m^2$ (the mass hierarchy) as has been discussed earlier \cite{IM}.

\begin{figure}[htp]
\includegraphics[width = \textwidth]{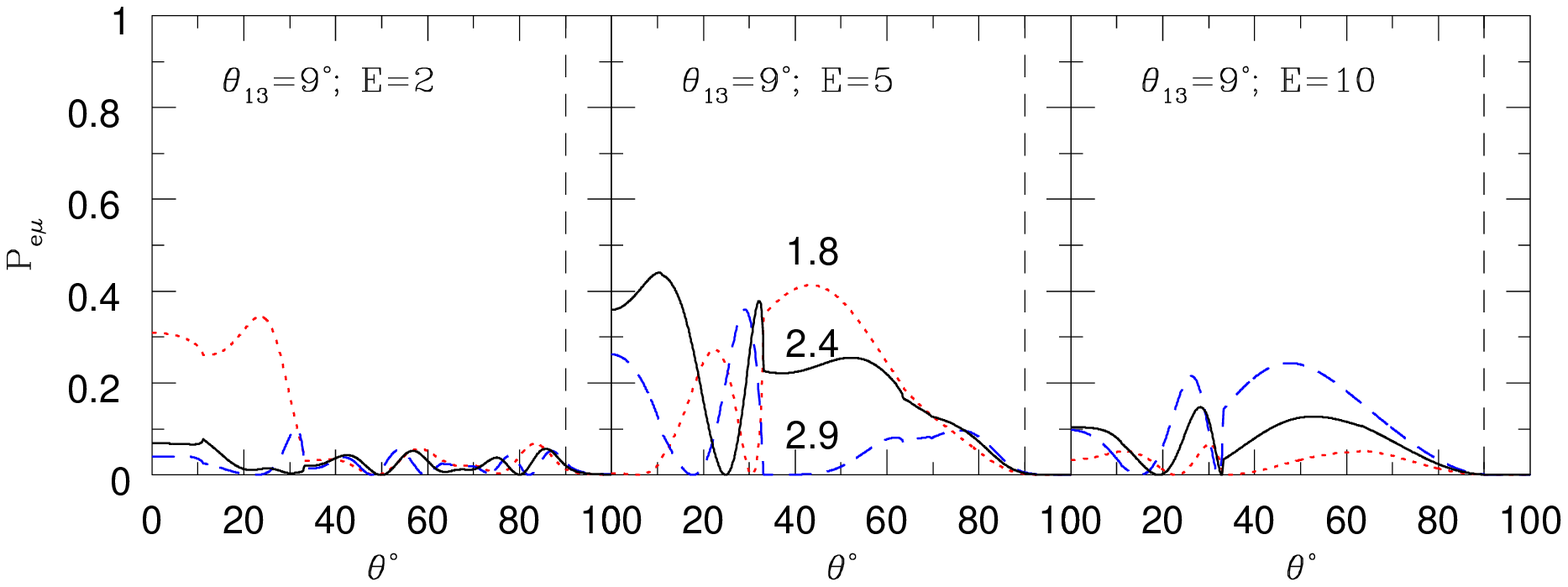}
\caption{Variations of $P_{e\mu}$ with $\Delta m^2$ at $\theta_{13} =
9^\circ$. See the caption to Fig.~\ref{fig:32t} for details.}
\label{fig:e32t}
\end{figure}

The conversion probability $P_{e\mu}$ is small unless $\theta_{13}$
is substantial. We show the high sensitivity of this probability
to variations in $\Delta m^2$ for $\theta_{13} = 9^\circ$ in
Fig.~\ref{fig:e32t}. An interesting feature is the substantially larger
probability for small nadir angles (0--$30^\circ$) and low energy ($E=2$
GeV) for the $\Delta m^2 = 1.8\times10^{-3}$ eV$^2$ value due to resonance
inside the core.  This feature is visible only at low $\Delta m^2$ values:
for example, for $E=2$ GeV, it occurs roughly for $\delta_{32} \sim \Delta
m^2 \lesssim 2.0 \times 10^{-3}$ eV$^2$, with a small uncertainty
in the limit due to the unknown $\theta_{13}$.  For larger values
of $\Delta m^2$, larger densities than are available inside the Earth
are needed to satisfy the resonance condition, Eq.~\ref{eq:resonance},
at $E=2$ GeV; hence there is no substantial enhancement of these
probabilities in the small nadir-angle region.

{\em A substantial ($> 20$\%) $P_{e\mu}$ in this region at small energies
around $E =2$ GeV is thus a clear indication that} $\Delta m^2 < 2
\times 10^{-3}$ eV$^2$. While this fact may not have an impact on the
overall event rates for atmospheric neutrinos, it will be important for
long baseline neutrinos since one can separately study $P_{e\mu}$ through
{\it wrong sign muons}. The anti-neutrino conversion probability, however,
remains small even for large $\theta_{13}$ and is therefore not shown.

\subsection{Variation of probabilities with $\theta_{23}$}

The Super-Kamiokande data favour a best-fit value of $\theta_{23} =
45^\circ$ in a 2-flavour analysis. Although a recent 3-flavour global
analysis \cite{foglipar} indicates that this parameter may be slightly
smaller, $\theta_{23} = 41.6^\circ$, the 2$\sigma$ allowed parameter
space spans both octants: $36^\circ \le \theta_{23} \le 52^\circ$. The
dominant (matter-independent) term in the survival probability depends
on $\sin^22\theta_{23}$. It is therefore difficult to substantially
decrease errors on $\theta_{23}$ via a measurement of the leading
contribution. Furthermore, even if a deviation from the maximal value
is measured, it is impossible to determine the octant with such a term.

Methods to determine the octant of $\theta_{23}$ have been discussed in
detail elsewhere \cite{probir}. Sensitivity to the octant of this angle
is realised through sub-leading terms that depend on $\theta_{23}$,
not $2\theta_{23}$, as can be seen from Eq.~\ref{eq:p3num}. These are
matter-dependent and sub-leading because they are also proportional to
$\sin\theta_{13}$, which is known to be small but is otherwise unknown.

\begin{figure}[htp]
\includegraphics[width = \textwidth]{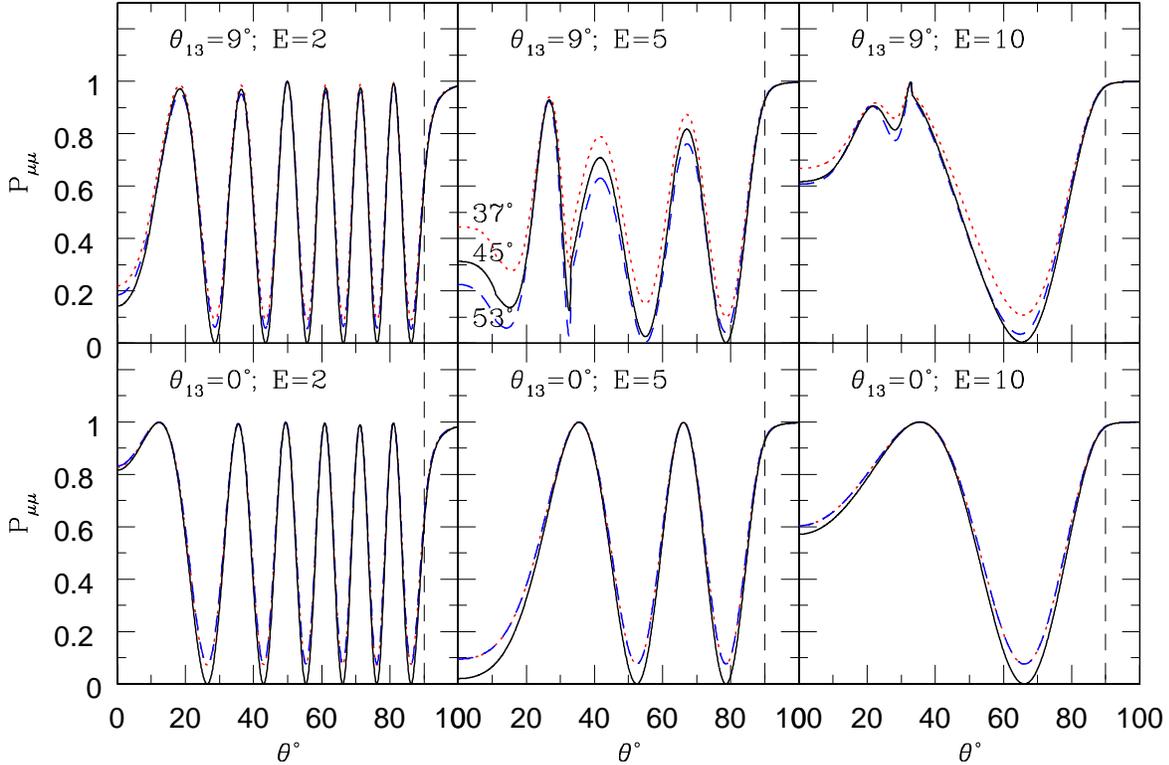}
\caption{Variation of $P_{\mu\mu}$ with $\theta_{23}$ for $\theta_{13} =
0,9^\circ$, as a function of the nadir angle. While probability curves
for $\theta_{23}$ values symmetrically larger and smaller than $45^\circ$
coincide when $\theta_{13}=0$ (bottom panels), they differ for non-zero
$\theta_{13}$ as shown in the top panels.}
\label{fig:23t}
\end{figure}

Fig.~\ref{fig:23t} shows the sensitivity of $P_{\mu\mu}$ to variations
in $\theta_{23}$. For clarity of discussion, we choose a central value
of $45^\circ$ (that is, choose a maximal value rather than the slightly
smaller central best-fit value as used earlier) and the {\em same}
2$\sigma$ variation about it. When $\theta_{13}=0$, symmetric 2$\sigma$
variations about the central value of $45^\circ$ lead to the {\em same}
survival probability, as can be seen at all neutrino energies in the
lower panels of the figure.

When $\theta_{13}$ is not zero, the survival probability for $\theta_{23}$
less than maximal (red dotted line) lies consistently above the
central value (solid line) for all values of $\Delta m^2$, $E$ and
nadir angles. This happens because the terms in $P_{\mu\mu}$ involving
$\sin\theta_{23}$ and $\sin 2\theta_{23}$ come with the same relative
(negative) sign, as seen from Eq.~\ref{eq:p3num}. As $\theta_{23}$
increases from the first octant towards maximal, both these terms are
increasing so that the probability systematically decreases. However,
as $\theta_{23}$ increases from the maximal to a value in the second
octant, $\sin \theta_{23}$ increases while $\sin 2\theta_{23}$ {\em
decreases}, leading to a complicated dependence of the probability on this
parameter. Hence no such regular behaviour is seen when $\theta_{23}$
is larger than $45^\circ$ although, in a large portion of phase space
at $E=5$--10 GeV, the probability for $\theta_{23} < (>)$ $45^\circ$
is systematically larger (smaller) than that at $\theta_{23} =45^\circ$.

It is clear from Fig.~\ref{fig:23t} that, for non-zero $\theta_{13}$,
the probability curve for $\theta_{23}$ in the first octant ($37^\circ$)
is better separated from that for $\theta_{23} = 45^\circ$, than the
curve for $\theta_{23}$ in the second octant ($53^\circ$). Evidently, the
ability to determine whether $\theta_{23}$ is different from maximal will
depend on which octant the true value of $\theta_{23}$ lies in. This is
also reflected in our numerical results with atmospheric neutrino events,
as discussed in the next section.

\begin{figure}[htp]
\includegraphics[width = \textwidth]{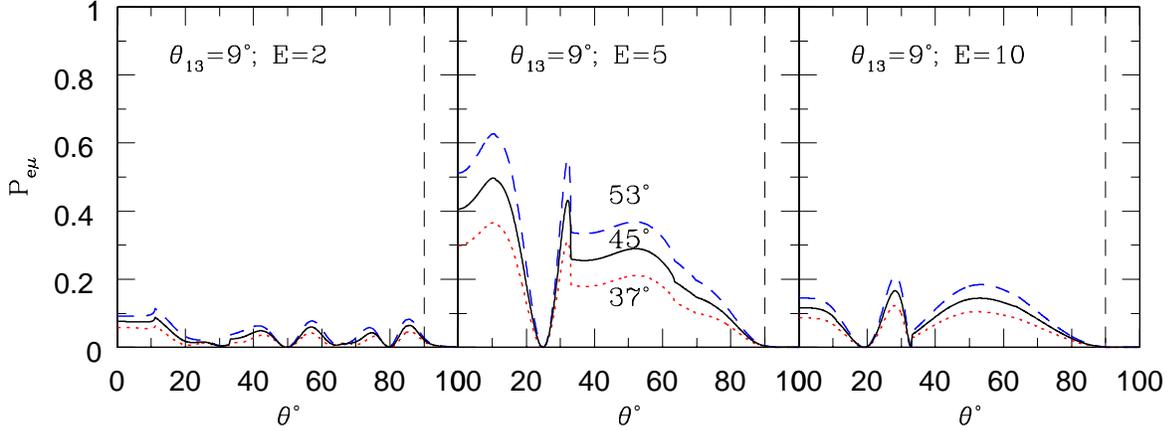}
\caption{Variation of $P_{e\mu}$ with $\theta_{23}$ for
$\theta_{13}=9^\circ$. }
\label{fig:e23t}
\end{figure}

Precisely the opposite effect is seen in the conversion probability
$P_{e\mu}$ as seen in Fig.~\ref{fig:e23t}. Here, the probability increases
with $\theta_{23}$ for all $\Delta m^2$ values at all energies and nadir
angles so that the probability for $\theta_{23} < (>)$ $45^\circ$ is {\em
systematically larger (smaller) than the probability with $\theta_{23}
= 45^\circ$.} This can be easily seen from the fact that there is a
single term proportional to $\sin\theta_{23}$ in Eq.~\ref{eq:p3num}
for $P_{e\mu}$, appearing with the opposite sign to the matter terms
in $P_{\mu\mu}$. Hence a long base-line experiment, where $\nu_e$ and
$\nu_\mu$ beams are separately available, will be able to clearly
establish the octant of $\theta_{23}$ by studying the $P_{e\mu}$
conversion probability via the wrong-sign muon events.

Note that if $\theta_{13} =0$ exactly, then the octant determination
cannot be made from $P_{\mu\mu}$ although $P_{e\mu}$ still shows the
same dependence on $\theta_{23}$ as discussed above. However, $P_{e\mu}$
is insignificant in this energy range and hence octant discrimination at
$\theta_{13}=0$ is possible only through a study of $P_{e\mu}$ at low
energies around $0.1$ GeV, when resonance effects in the (12) sector
enhance this probability. This is outside the scope of the present
paper and will be discussed in a separate publication.

This relatively systematic dependence on $\theta_{23}$ is in contrast to
the rather complex dependence of the survival and conversion probabilities
on $\Delta m^2$ and $\theta_{13}$.  Both these probabilities contribute
to the observed muon events in studies of atmospheric neutrinos.
We shall explore this further in a later section to maximise this
``octant-seeking'' effect.

In contrast, the matter dependence is small for the anti-neutrinos;
hence, the anti-neutrino survival probabilities for all $\theta_{13}$
values are similar to the neutrino survival probability at $\theta_{13}
=0$ and do not distinguish the octant of $\theta_{23}$. We reiterate
that this is a consequence of the chosen hierarchy; opposite results
are obtained with the inverted hierarchy (matter effects are ultimately
larger in the normal hierarchy since the anti-neutrino cross-sections
are smaller than the neutrino cross-sections so that event rates are
larger and matter effects better differentiated in the neutrino sector).

\subsection{Variation of the probabilities with Earth density}

In order to compute the effect of Earth's matter on the probabilities,
we have used the {\sc prem} model \cite{prem} which gives the density
profile of a ``spherical equivalent Earth''. While the locations of the
discontinuities and mantle-core transitions are rather well-known,
the absolute values of the densities themselves are not so well
established. Hence, in order to study effects of uncertainties in the
Earth's density, we have retained the locations of the discontinuities
while changing the density values. We have done this in a simple manner by
changing all the densities within the core by $\Delta \rho_{\rm core}$
= 0.6 gm/cc (a change of roughly 5\%, as allowed by the model), and
suitably adjusting the mantle densities in such a way as to maintain the
overall mass of the Earth to be constant. Such a procedure results in a
much smaller change in the mantle densities of $\Delta \rho_{\rm mantle}
\sim - 0.2 \Delta \rho_{\rm core}$, in the opposite direction.

\begin{figure}[htp]
\includegraphics[width = \textwidth]{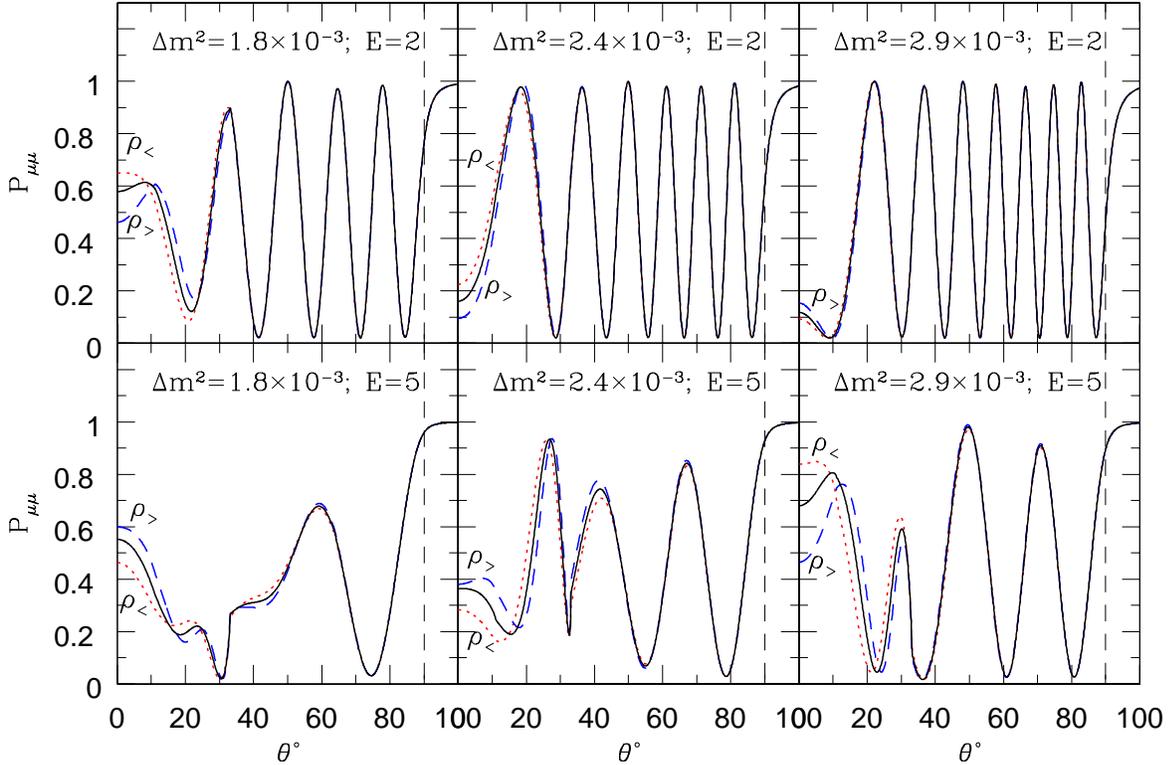}
\caption{Variation of ${P}_{\mu\mu}$ with changes in the Earth's density
for $\theta_{13} = 9^\circ$ and a 2$\sigma$ variation in $\Delta m^2$
at $E=2,5$ GeV.  The curves labelled $\rho_>$ and $\rho_<$ correspond
to increasing and decreasing the core densities by roughly 5\% (with a
compensating change in the mantle densities) compared to the black solid
line for the {\sc prem} density profile.}
\label{fig:det}
\end{figure}

\begin{figure}[htp]
\includegraphics[width = \textwidth]{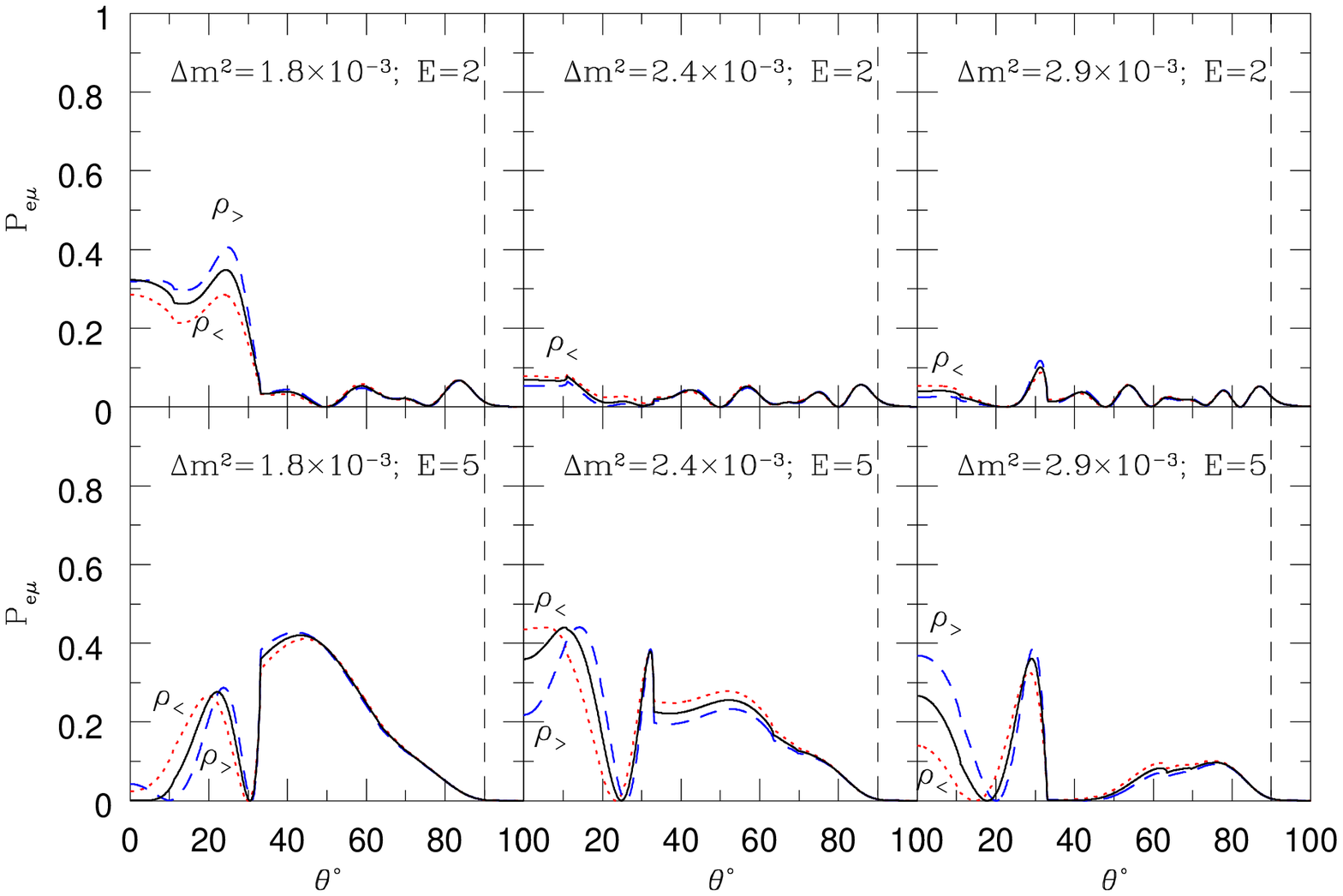}
\caption{As in Fig.~\ref{fig:det} for $P_{e\mu}$.}
\label{fig:edet}
\end{figure}

The consequences of such a density change to the neutrino survival
and conversion probabilities are shown in Figs.~\ref{fig:det} and
\ref{fig:edet} respectively. Here $\rho_>$ and $\rho_<$ correspond to
increasing and decreasing the core densities by 5\%, with the mantle
densities adjusted suitably. The solid black line is for the original
{\sc prem} density profile. Again, the size of matter effects depends
on $\Delta m^2$; changing the density alters the $\Delta m^2$ at which
resonance occurs, for a given neutrino energy.

A point to note is the difference between the {\sc prem} profile and
constant density slabs. The former has a continuously changing density
profile, even within slabs. Hence neutrinos with a band of energy can
undergo resonance in any given slab. In contrast, only a neutrino with
a single definite energy can undergo resonance in a constant density
slab (apart from having purely adiabatic propagation inside any slab). Due
to this, the probability curve as a function of zenith angle for a
given energy can look very different when computed with the {\sc prem}
profile and with a constant density approximation, even when the latter
closely follows the {\sc prem} profile. However, when an integration
over an energy bin is performed, such variations get averaged
out. More details on this issue are given in the Appendix.

At low energies where the matter effects are significant only for
core-crossing neutrinos at small nadir angles, the effects of altering
the density profile are also seen only at these small nadir angles, as
can be seen from the upper panels of the figures. At higher energies,
changes appear at larger nadir angles as well, as can be seen from the
lower panels, since resonance (and enhanced matter effects) occurs in
the mantle itself.  While the variations are small, except at small
nadir angles, again, they are complex and depend on both energy and a
precise knowledge of $\Delta m^2$. Probing the Earth density, especially
in the core, with neutrino oscillation physics using neutrino factory
beams has been discussed earlier \cite{density}. Here, only the dependence
on $\theta_{13}$ was considered while $\Delta m^2$ was kept fixed.

We remark that in all cases there is also a dependence on $\theta_{13}$
at small nadir angles since the variations with $\theta_{13}$ are
substantial here, as can be seen from Fig.~\ref{fig:13t}. The dependence
on density variations may be disentangled from that on $\theta_{13}$
since substantial variations in the intermediate nadir angle region can
only be due to $\theta_{13}$, as seen from Fig.~\ref{fig:13t}. 

The anti-neutrino probabilities again do not show any significant
dependence on these parameters in the normal hierarchy.

\subsection{Variation of probability with the CP phase}

\begin{figure}[htp]
\includegraphics[width = \textwidth]{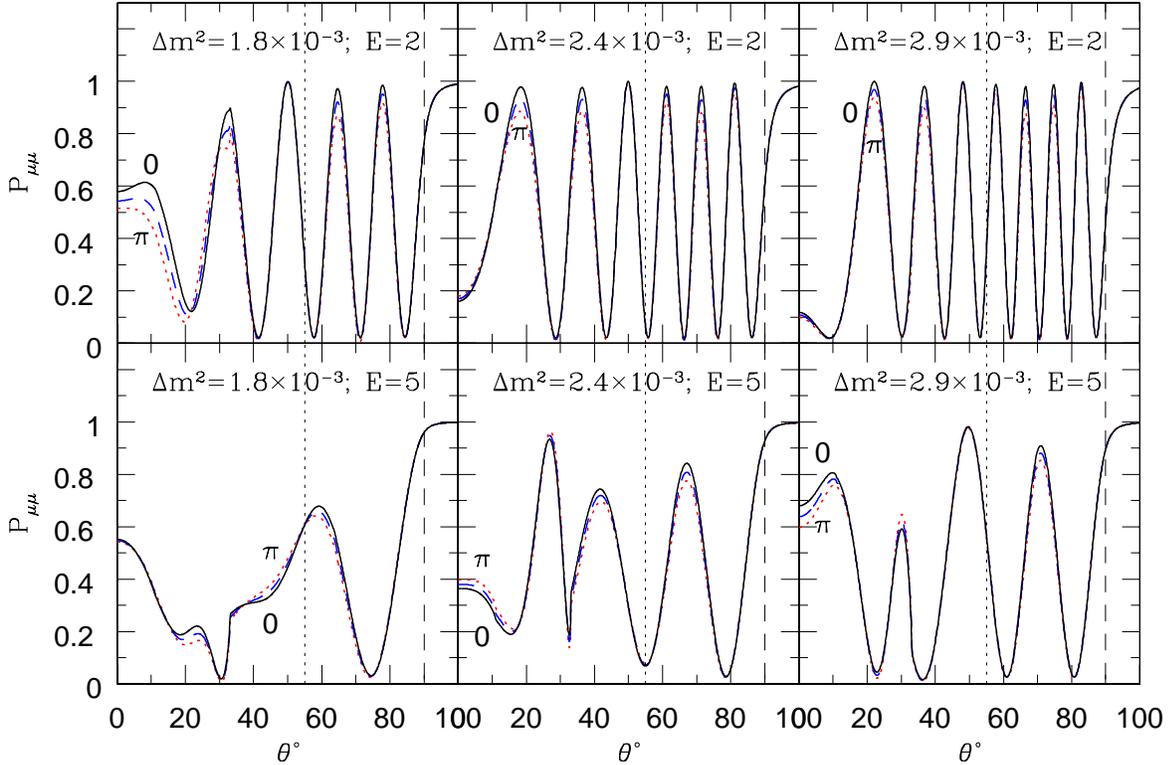}
\caption{Variation of $P_{\mu\mu}$ as a function of nadir angle with the
CP phase $\delta$ for $\theta_{13} = 9^\circ$, for a 2$\sigma$ variation
of $\Delta m^2$. The curves shown are for $\delta = 0, \pm\pi/2$, and
$\pm\pi$. The vertical dotted line indicates the magic base-line where
the probabilities are independent of $\delta$.}
\label{fig:cpt}
\end{figure}

We now explore the dependence of the probabilities on the CP phase
$\delta$. This dependence is shown in Fig.~\ref{fig:cpt}, again
for $\theta_{13}=9^\circ$. Apart from the dashed vertical line at
$\theta=90^\circ$ that separates the ``up-going'' probability from the
``down-going'' one, the dotted vertical line at $\theta \sim 55^\circ$
indicates the nadir angle corresponding to the so-called ``magic
base-line'' \cite{magic} where the probabilities are independent of
the CP phase. Since the Earth mantle density is not a constant, this is
actually a band around that nadir angle, as can be seen from the figure.

The effect of the CP phase $\delta$ competes with the matter effect which
in turn depends on the relative size of the matter term $A$ compared
to $\Delta m^2$. The variation of the probabilities (as a function of
energy and nadir angle) with $\delta$ for the current 2$\sigma$ allowed
range of $\Delta m^2$ is shown in Fig.~\ref{fig:cpt}. This variation is
so small that it is unlikely to be measured via a study of atmospheric
neutrinos and may need long base-line measurements where the spectrum
and nadir angle are well-known. Nevertheless, we make a few remarks on
the general behaviour.

While the survival probability $P_{\mu\mu}$ is symmetric for $\delta =
\pm \pi, \pm \pi/2$, $P_{e\mu}$ can distinguish the cases $\delta = \pm
\pi/2$. At low energy, $E=2$ GeV, the top panels in Fig.~\ref{fig:cpt}
indicate that the survival probability $P_{\mu\mu}$ is typically larger
for $\delta=0$ at all nadir angles, smaller for $\delta = \pm\pi/2$ and
smallest for $\delta =\pm\pi$, {\em independent} of $\Delta m^2$. This
is not true at larger energies, as can be seen from the lower panels of
the figure. Hence, if these small variations can indeed be measured,
information on $\delta$ may be obtained with some confidence, only if
$\Delta m^2$ is known to much better than the current precision.

\begin{figure}[htp]
\includegraphics[width = \textwidth]{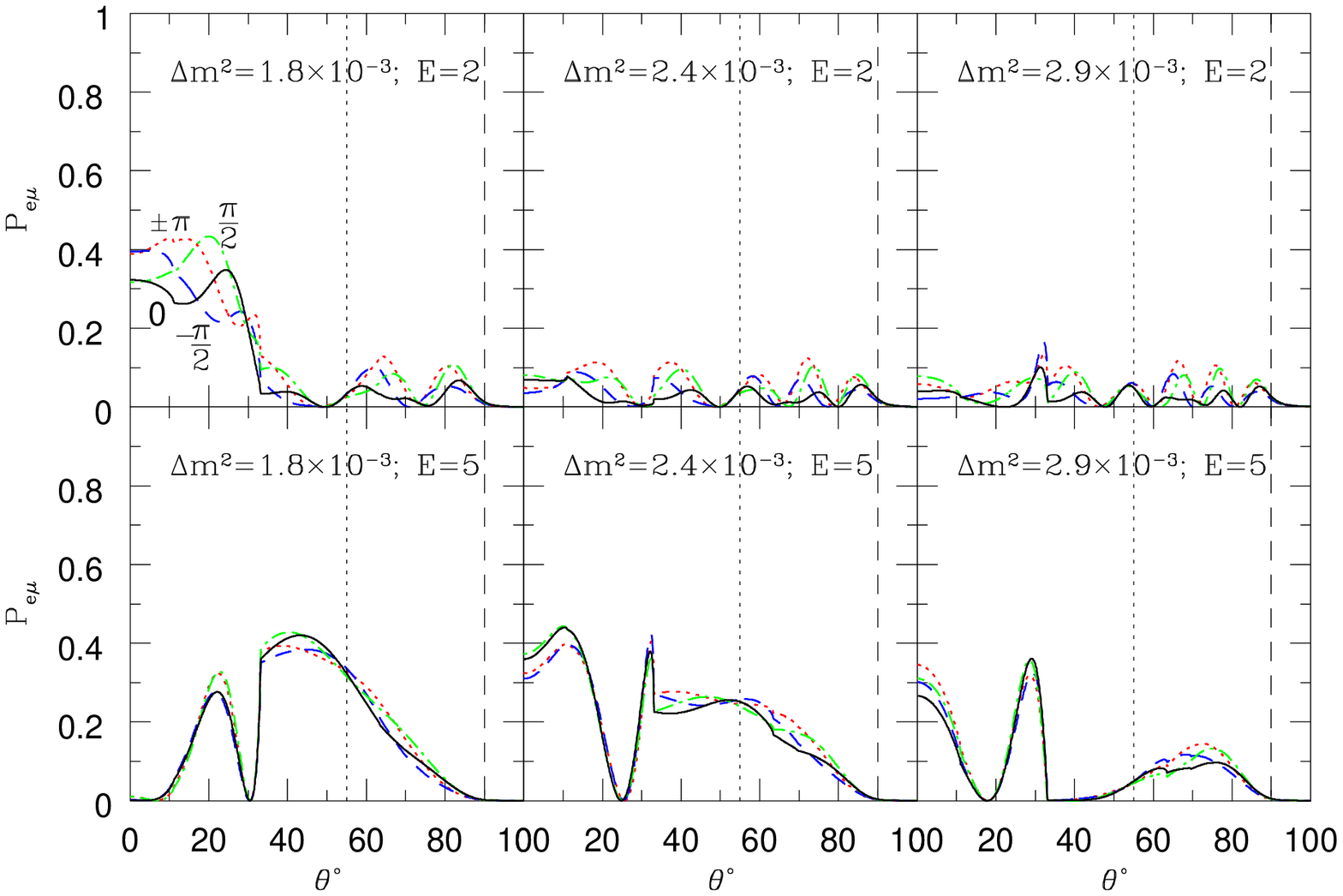}
\caption{As in Fig.~\ref{fig:cpt} for $P_{e\mu}$. The curves shown are
for $\delta = 0, \pi/2, -\pi/2$, and $\pm\pi$.} 
\label{fig:ecpt}
\end{figure}

The situation is murkier with the conversion probability $P_{e\mu}$, as
can be seen from Fig.~\ref{fig:ecpt}. The clear trend that was visible
with $P_{\mu\mu}$ at low energy is no longer present. At $E=5$ GeV there
is a systematic trend between the curves for $\delta = 0$ and $\pi/2$
and between $\delta = -\pi/2$ and  $\pm \pi$; they behave in the same
way relative to each other. The complex effect of $\delta$ is overlaid
on an already small $P_{e\mu}$.  Signatures of $P_{e\mu}$ are very clean
however with long-base-line detectors (via wrong-sign muon measurements);
this may facilitate a measurement of $\delta$ at such detectors, as has
been discussed elsewhere \cite{cp}.

The anti-neutrino probabilities are rather insensitive to matter effects
or the CP phase (in the normal hierarchy) and we do not discuss them
here.

\section{Event rates in a detector}

\subsection{Preliminaries}

While one can identify clearly the regions of sensitivity to the
fundamental neutrino parameters through an analysis of the oscillation
probabilities, it is crucial to identify the appropriate observables which
will enable a precision measurement of the parameters. The experimental
observables are primarily the event rates in a detector. The fully
differential event rate for neutrinos of flavour $\alpha$ to be detected
is given by the general expression:
\begin{equation}
\frac{\d N^{\alpha}}{\d E\d x} = 
K_y\sum_{\beta} \Phi_\beta(E, x) ~ P_{\beta\alpha}(E,x)
~ \sigma_{\alpha}(E)~,
\label{eventr}
\end{equation} 
where $x = L/E$ and $\sigma_{\alpha}$ is the total interaction
cross-section for the $\alpha$ type neutrino to interact with the detector
material. Here $P_{\beta\alpha}$ is the conversion probability of a
neutrino of flavour $\beta$ to a flavour $\alpha$ (which also includes
the survival probability when $\alpha=\beta$).

The flux-dependent term $\Phi_\beta (E, x)$ is related to the
doubly-differential neutrino (or anti-neutrino) flux of flavour
$\beta$, $\d^2 \phi_{\beta}(E,\theta)/\d \ln E \, \d \cos\theta$, which
is a function of the energy $E$ and nadir angle $\theta$ through a
Jacobean of transformation.  In the case of atmospheric neutrinos a
range of zenith angles are available while for long-baseline neutrinos
this angle is kept fixed by the location of the source.

The factor $K_y$ is the detector dependent factor measured in units
of kton-years. While we assume the detector to be mainly made up of
magnetised iron with active detector elements, this is not crucial since
only the detector mass and exposure times are relevant. In the {\sc
monolith} and {\sc ical/ino} proposals, where muons from
charged-current $\nu_\mu$-nucleus interactions are detected, the active
detector elements are resistive plate chambers (RPCs, gas-filled glass
chambers). In either of these proposals the detector mass is almost
entirely ($>$ 98\%) due to its iron content. We will be interested here
in {\em event ratios}; hence the factor $K_y$ and other actual detector
details will only determine the errors. Furthermore, we assume that the
detector is capable of correctly identifying the charge of the muons
in the final state. This will be better than 98\% for energies $2 \le
E \le 10$ GeV for the {\sc ical/ino} detector. This is crucial for
the precision determination of the neutrino parameters in any future
experiment.

\subsection{Atmospheric Neutrinos}

We now focus on neutrino oscillation studies with atmospheric
neutrinos. Both $\nu_\mu$ and $\nu_e$ (and their anti-particles) are
produced, typically in the ratio 2:1. 

The distance of propagation $L$ of the neutrino from the point of
production to the detector is given by,
\begin{equation}
L = \sqrt{(R_0+L_0)^2-(R\sin\theta)^2} +  R\cos\theta~,
\label{Ldef}
\end{equation}
where $L_0$ is the average height (taken to be 15 km) above the surface
of the Earth at which the atmospheric neutrinos are produced, $R_0$ is
the radius of the Earth and $R=R_0-d$, $d$ being the depth at which the
detector is located underground (chosen to be 1 km). Note that $\theta=0$
corresponds to neutrinos reaching the detector vertically upwards.

The event rate in a given bin of $x = L/E$ is,
\begin{equation}
N^{\alpha}_{\rm bin}(x) = \int_{\rm bin}  \d x \int_{E_{\rm min}} \by{\d E}{E}
\frac{\d^2 N^{\alpha}}{\d\ln E\d x}~;
\label{eventb}
\end{equation}
henceforth we discuss only the case of muon-neutrinos, $\alpha=
\mu$ (or $\overline{\mu}$). Then both $P_{e\mu}$ and $P_{\mu\mu}$
contribute in the expression above. However, due to the smallness of
the (13) mixing angle, $\theta_{13}$, the contribution of $P_{e\mu}$
is generally small and is significant only near large $L/E$ where it is
about 10--20\% (5\%) for neutrinos (anti-neutrinos).

The event rate is expressed as a function of $x$, averaged over a bin
width that will be appropriately chosen to maximise the sensitivity to
the sign of $\delta_{32}$ or the octant of $\theta_{23}$. The expression
given in Eq.~\ref{eventb} is the best-case scenario since the integration
is over the neutrino or anti-neutrino energy. We discuss the effect
of including the detector resolution functions in the next section and
proceed now to analyse the ideal case with a perfect detector.

\subsection{Event rates with an ideal detector}

\begin{figure}[htp]
\includegraphics[width = 5cm]{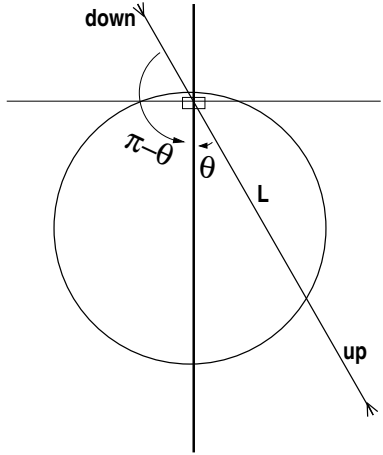}
\caption{A schematic showing the up- and down- neutrino directions and
their labelling in the nadir angle $\theta$.}
\label{fig:updown}
\end{figure}

A useful measure of oscillations is the ratio of up-coming to down-going
neutrinos with nadir/zenith angles interchanged. This is clear from
Fig.~\ref{fig:updown}. The fluxes of atmospheric neutrinos from directions
$\theta$ and $(\pi - \theta)$ are expected to be similar in the absence
of oscillations, especially for larger energies ($E$ greater than a few
GeV). Since the path-length traversed, $L$, is related to $\theta$ as
$$
L = f(\vert \cos\theta \vert) + R \cos\theta~,
$$
(see Eq.~\ref{Ldef}), the replacement $\theta \leftrightarrow (\pi -
\theta)$ effectively changes the sign of the second term in the equation
above, thus taking, for instance, a down-going neutrino to an up-coming
one.
The ratio of events in the up-down directions for a given $x =
L/E$, therefore, reflects the asymmetry of the up-down fluxes, due to
oscillations. We define \cite{picchi},
$$
{\cal{R}} = \by{U}{D}(x) =
\by{\hbox{No. of events from up-coming muon neutrinos} (x)}
{\hbox{No. of events from down-going muon neutrinos} (\tilde{x})}~,
$$
where $\tilde{x} = x (\theta \leftrightarrow (\pi - \theta))$ and the
number of up-coming (U) or down-going (D) events is calculated using
Eq.~\ref{eventb}. Similarly the ratio $\overline{U}/\overline{D}$
defines the corresponding up/down ratio for anti-neutrinos. Since the
effect of oscillations on the denominator is small, the ratio ${\cal{R}}$
is effectively the ratio of oscillated to unoscillated events with the
same $L/E$.

While reflecting the effect of oscillations, this ratio also minimises
errors due to the uncertainties in the overall flux normalisation
(which can be as large as 30\%) and those in the cross-sections (about
10\%). We now proceed to determine these ratios numerically, and study
their sensitivity to the oscillation parameters of interest.

We use the 3D atmospheric neutrino fluxes as calculated by Honda et al.
\cite{honda}. These have been obtained from the {\sc nuance} neutrino
generator \cite{nuance}. The charged current muon-neutrino nucleus
cross-sections used have also been extracted from the {\sc nuance} code.
At the low end of the energy spectrum, the interaction with the detector
material is mainly through quasi-elastic and resonance interactions. When
the energy is a few GeV, deep inelastic scattering (DIS) dominates,
ultimately taking over by about 10 GeV, as this cross-section is
proportional to the neutrino energy. There are still issues about the
transition from the resonance (mainly single pion dominated) region
to the DIS region \cite{nuance}; however, such errors are minimised
by {\em always analysing ratios of events}. Roughly speaking, in the
energy range of interest from few to 10's of GeV, half the events are from
quasi-elastic and resonance interactions and the other half from DIS. This
proportion is somewhat sensitive to finite detector resolutions, since the
neutrino energy is smeared over a much larger range than that observed;
however, the events are still dominantly low-energy ones owing to the
steep fall (faster than $1/E^2$) of the incident atmospheric neutrino
flux with energy.

We find, as studies of the probabilities suggest, that variations of
the event rates due to uncertainties in $\delta_{21}$ and $\theta_{12}$
are small. We therefore fix them to their best-fit values. We also set
the CP phase $\delta=0$ and fix the Earth's density to be as given by the
{\sc prem} density profile \cite{prem}. The remaining parameters, viz.,
$\Delta m^2$ (both magnitude and sign), $\theta_{13}$ and $\theta_{23}$
(in particular, its deviation from maximality and hence also its octant),
are the parameters to be determined or constrained in the analysis. A
study of atmospheric neutrino events is sensitive to all these parameters
(but the magnitude of $\Delta m^2$) only through matter effects. That
is, $\theta_{13}$ must be non-zero and substantial for there to be any
sensitivity to these parameters. Since this parameter is known to be
small, it therefore follows that large exposures are required to determine
any of the above parameters to better precision than are currently known.

In order to proceed further, we shall assume in the following that
$\theta_{13}$ is substantial; the precise threshold will be determined
through the analysis. It is already known that the sign of $\Delta m^2$
can be determined via the {\em difference asymmetry} of the up/down
ratios in the neutrino and anti-neutrino channels:
\begin{equation}
{\cal{A}} = U/D - \overline{U}/\overline{D}~, 
\label{eq:da}
\end{equation}
where $U$ and $D$ are the relevant differential event rates for up and
down events as a function of $L/E$ and $\overline{U}$ and $\overline{D}$
the corresponding rates for anti-neutrinos. This asymmetry can be maximised
by choosing to integrate over $L/E$ bins that include one half-period
of the oscillation. This can be approximated by the matter-independent
condition,
\begin{equation}
\by{1.267\vert \delta_{32}\vert L}{E} = n \by{\pi}{4}~; \quad n \hbox{ odd}.
\label{eq:asymbins}
\end{equation}
A detailed analysis of this asymmetry has already been done in
Ref.~\cite{IM}; we only remark here that we require $\sin^22\theta_{13}
\ge 0.06$ ($\theta_{13} \ge 7^\circ$) for the sign of $\Delta m^2$ to be
determined using this technique. Furthermore, this determination requires
750--1000 kton-year exposure, depending on the value of $\theta_{13}$,
thus setting the scale for our present analysis.

We focus therefore on the precision to which the magnitude of $\Delta
m^2$ and the octant of $\theta_{23}$ can be determined, given such
large exposures. (Current data from the Super-Kamiokande collaboration
restrict these to the ranges $1.8 \le \vert \Delta m^2 \vert \times 10^3$
eV$^2 \le 2.9$ and $36.5 \le \theta_{23} \le 54.5^\circ$.) This again
depends on $\sin^2 \theta_{13}$, which is expected to be shortly measured
with good precision (to within a few percent) by the Double-{\sc chooz}
experiment \cite{dchooz}. We study both cases: when $\theta_{13}$ is
known and hence kept fixed in the analysis, as well as the case when it
is allowed to vary freely. We restrict ourselves to the normal hierarchy,
with $\Delta m^2$ positive, so that matter effects (and hence sensitivity
to the octant of $\theta_{23}$), are enhanced in the neutrino channel. We
integrate event rates over an energy range of $E = 5$--10 GeV to maximise
statistics while retaining sensitivity to matter effects.

\begin{figure}[htp]
\includegraphics[width = \textwidth]{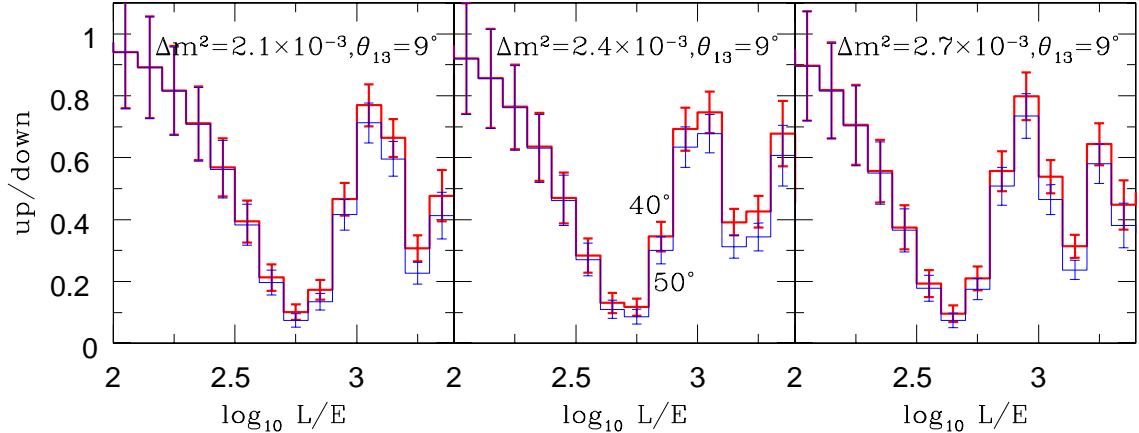}
\caption{Variation of the ratio of the rates for up-coming and down-going
neutrinos, integrated from $E=5$--10 GeV, in bins of $\log_{10}L
\hbox{ (km)}/E \hbox{ (GeV)} = 0.1$.
The variation with $\Delta m^2$ is shown in the three panels,
where results for a $\Delta m^2$ over the currently allowed 1$\sigma$
range, via., $\Delta m^2 = (2.1, 2.4, 2.7) \times 10^{-3}$ eV$^2$ are
shown. The two histograms in each panel correspond to $\theta_{23} =
40^\circ$ (upper) and $50^\circ$ (lower). The value of $\theta_{13}$
is kept fixed at $9^\circ$. Statistical error bars corresponding to
an exposure of 1000 kton-years are also shown.}
\label{fig:deltam223}
\end{figure}

In Fig.~\ref{fig:deltam223} we show the variation of the up/down events
ratio for different $\Delta m^2$ values as a function of $\log_{10}L/E$
for two values of $\theta_{23} = 40^\circ, 50^\circ$, in two octants. Here
$\theta_{13}$ is fixed to $9^\circ$. Two features are immediately
noticeable: (1) the position of the minima and maxima in $L/E$ are
not altered by changing $\theta_{23}$; furthermore, the events ratio
at the first minimum is not very sensitive to $\theta_{23}$, and (2)
the effect of changing $\theta_{23}$ from a value in the first octant
to a corresponding one in the second octant is to systematically {\em
decrease the event rates in all bins} for all $\Delta m^2$ values.

Furthermore, although not shown in the figure due to constraints of
clarity, the curve for maximal $\theta_{23} = 45^\circ$ lies between the
two curves for $\theta_{23} = 40, 50^\circ$ in all $L/E$ bins beyond the
first minimum. Such a systematic decrease with increase in $\theta_{23}$
was seen only in some zenith angle and energy ranges (for example for
$\theta = 0$--$70^\circ$ for $E=5$ GeV and for $\theta = 0$--$50^\circ$
for $E=10$ GeV, when $\theta_{13}=9^\circ$; see the top panels of
Fig.~\ref{fig:23t}).  By a judicious choice of the energy and $L/E$
interval, this effect has been converted to a systematic difference for
all bins to the right of the first minimum; moreover, it is in these
bins that the sensitivity to $\theta_{23}$ is significant, so that this
dependence is robust and easy to observe.

\begin{figure}[htp]
\includegraphics[width = \textwidth]{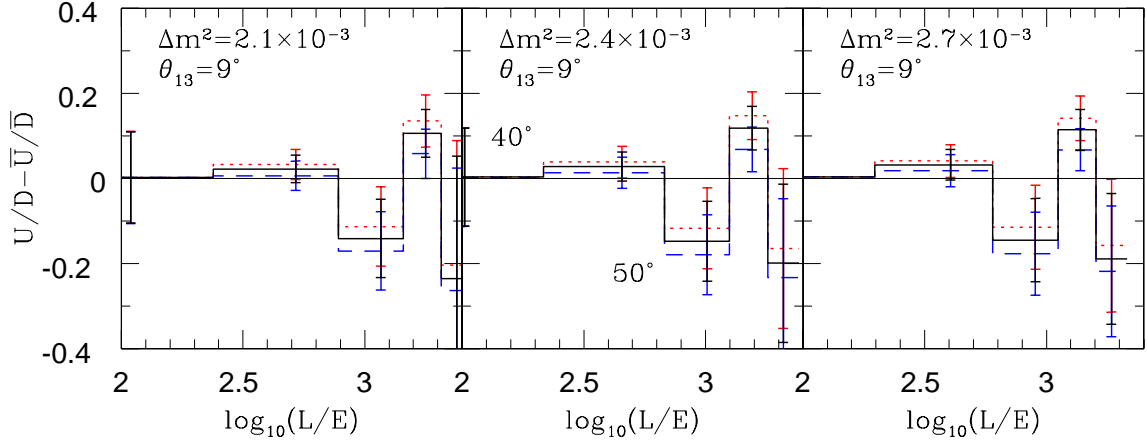}
\caption{As in Fig.~\ref{fig:deltam223}, but for the difference asymmetry,
$A = U/D - \overline{U}/\overline{D}$. Three histograms are shown, for
$\theta_{23} = 40^\circ$ (dotted), $45^\circ$ (solid) and $50^\circ$
(dashed lines), as a function of $\log_{10} L/E$ in bins defined in
Eq.~\ref{eq:asymbins}. It is seen that the histograms for $\theta_{23}
< 45^\circ$ and $\theta_{23} > 45^\circ$ are systematically greater and
less than that for $\theta_{23} = 45^\circ$.}
\label{fig:lebin}
\end{figure}

Note that the difference asymmetry defined in Eq.~\ref{eq:da} is
a difference between neutrino and anti-neutrino rates ratios. In this
case, the systematic decrease with increasing $\theta_{23}$ is true in
{\em all} $L/E$ bins for any $\Delta m^2$ value, by definition. That is,
the result for $\theta_{23} = 45^\circ$ always lies between that for a
value of $\theta_{23}$ in the first and second octant. This is shown
in Fig.~\ref{fig:lebin}. However, as can be seen from the relative
size of the error bars, the task of actually extracting the octant of
$\theta_{23}$ from measurement of such an asymmetry is severely limited
by statistics.  In fact, it is the poor statistics in the anti-neutrino
sector that limits the efficacy of this parameter.

\begin{figure}[htp]
\includegraphics[width = \textwidth]{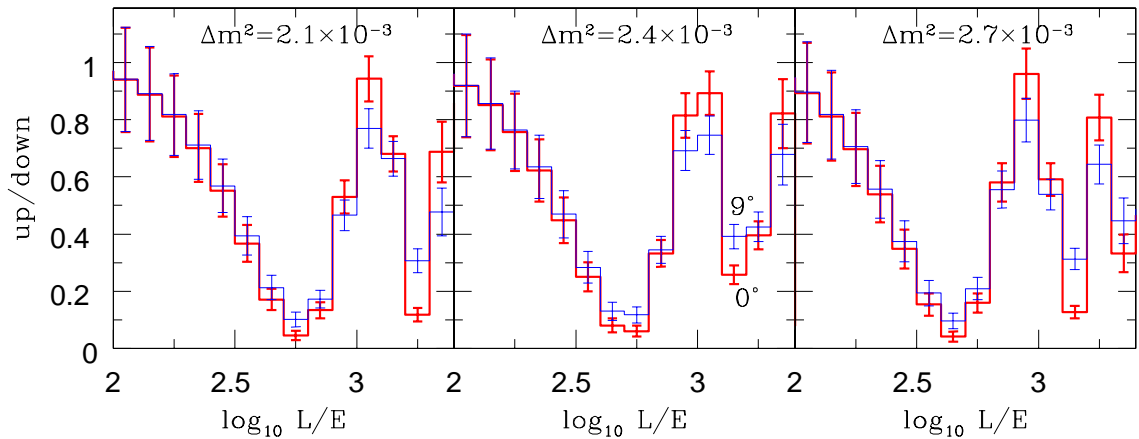}
\caption{As in Fig.~\ref{fig:deltam223}, but for fixed $\theta_{23} =
40^\circ$, while $\theta_{13} = 0,9^\circ$ in each panel. The outer
histogram (larger at maximum and smaller at minimum) corresponds to
$\theta_{13}=0$.}
\label{fig:deltam213}
\end{figure}

The effect of increasing $\theta_{13}$ away from zero is seen in
Fig.~\ref{fig:deltam213} where the events ratio is plotted as a function
of $\log_{10} L/E$ for different $\Delta m^2$ values. The two
histograms for each $\Delta m^2$ value correspond to $\theta_{13} =
0,9^\circ$, where the latter value is simply chosen as being within the
range allowed by current data and where matter effects are known to be
substantial. The effect of increasing $\theta_{13}$ is to turn on matter
effects that move the muon survival probability maxima (as a function of
$\log_{10} L/E$) away from 1 and the minima away from zero. This means
that for non-zero $\theta_{13}$ the events ratio is smaller at probability
maxima and larger at probability minima than for $\theta_{13} =0$. (Note
that the extent of deviation is somewhat suppressed by the sub-dominant
contribution of $P_{e\mu}$ to the dominant $P_{\mu\mu}$ terms; experiments
such as those with neutrino factories where such contamination does not
exist will be even more sensitive to matter effects).

The contrasting behaviour of the events up/down ratio with $\theta_{13}$
and $\theta_{23}$ for a given $\Delta m^2$ is clear: an increase in
the latter systematically decreases the ratio, while an increase in the
former decreases the ratio at the $\log_{10} L/E$ maxima and increases
them at the $\log_{10} L/E$ minima. Such behaviour enables
the extraction of the different parameters when the data is fitted to
the theory.

To demonstrate this (and to determine the efficacy and precision with
which the parameters can be determined), we begin by computing up/down
neutrino events ratios for the set of input parameters, $(\theta_{13},
\theta_{23}, \Delta m^2) = (7^\circ, 40^\circ, 2.4\times 10^{-3}$
eV$^2)$, in different $\log_{10} L/E ({\rm km/GeV})$ bins from 1.6
to 3.4 of width 0.1 each. We then fit this ``data'' to these three
parameters, and extract the allowed parameter space to establish how
well these parameters can be determined. In particular, we focus on
the precision of measurement of all parameters as well as the possible
determination of the octant of $\theta_{23}$. Since we are considering
ratios, we include only statistical errors in our analysis. Note that
a zenith angle cut on very horizontal events that may not be easily
detected by horizontally aligned detectors such as {\sc ical/ino} only
removes the small $L/E$ bins. For instance, a cut of $\cos\theta = 0.1$
leads to a cut on $\log_{10} L/E \ge 2.15$. For such small values, as can
be seen from Fig.~\ref{fig:deltam213}, there is negligible dependence
of the up/down ratio on the neutrino oscillation parameters. Hence,
such cuts will not affect our analysis.

\begin{figure}[htp]
\includegraphics[width = \textwidth]{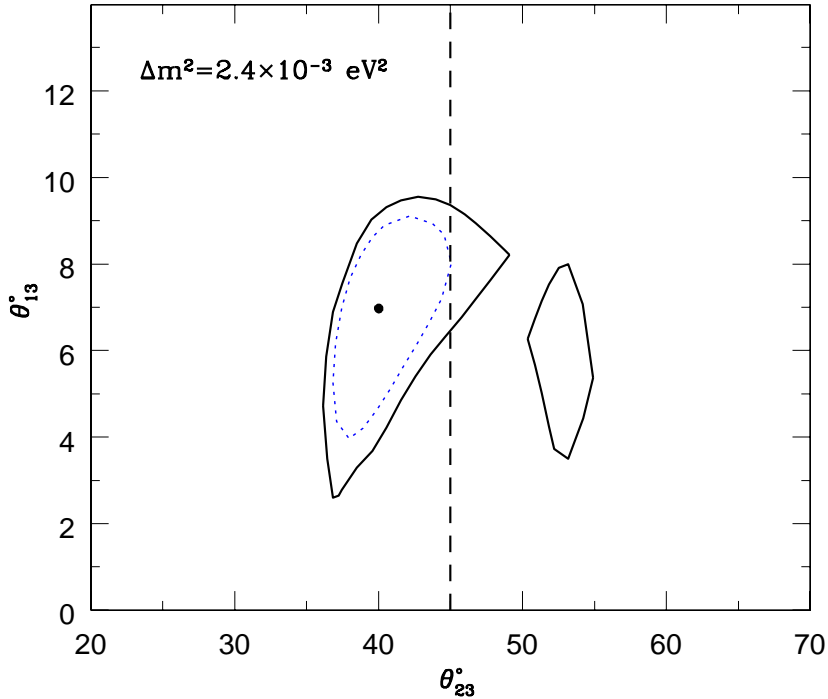}
\caption{Allowed parameter space in $(\theta_{23}$,$\theta_{13})$
variables. Input used was $(40^\circ, 7^\circ)$, for $\Delta m^2 =
2.4\times 10^{-3}$ eV$^2$ (shown as the solid point, also the
best-fit). Shown are the 95\% and 99\% CL contours; it is seen that the
``wrong-octant'' solution to the right of $\theta_{23} = 50^\circ$ is
allowed at 99\% CL.}
\label{fig:m37}
\end{figure}

Fig.~\ref{fig:m37} shows the allowed region at 95\% CL and 99\% CL in the
$(\theta_{13}, \theta_{23})$ parameter space, using standard $\chi^2$
minimisation, for a fixed value of $\Delta m^2 = 2.4\times 10^{-3}$
eV$^2$. At 95\% CL, the up/down ratio barely disallows maximal mixing
in $\theta_{23}$; it also disallows the ``wrong'' octant solution of
$\theta_{23}= (\pi/2 - \theta_{23}^{\rm input}) = 50^\circ$. However,
both these sensitivities go away at the 99\% CL level when an island of
allowed parameter space opens up around the wrong-octant solution, in a
region to the right of it.

\begin{figure}[htp]
\includegraphics[width = \textwidth]{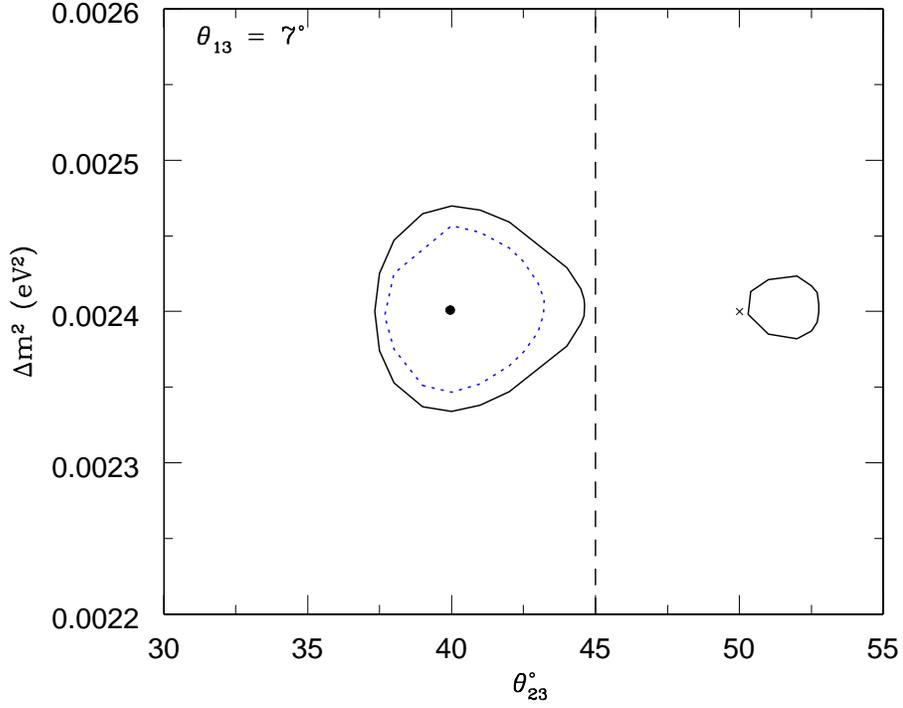}
\caption{Allowed parameter space in $(\theta_{23}, \Delta m^2)$
variables. Input used was neutrino up/down event rates at
$(40^\circ, 2.4\times 10^{-3}$ eV$^2$), for $\theta_{13} = 7^\circ$.
Shown are the 95\% and 99\% CL contours.}
\label{fig:mdelta37}
\end{figure}

At this point we note that $\theta_{13}$ is likely to be rather
precisely fixed (to much better than a percent) by the Double-{\sc
chooz} experiment \cite{dchooz}. We therefore observe the effect of
keeping $\theta_{13}$ fixed and varying the other two parameters. This
is shown in Fig.~\ref{fig:mdelta37} where the allowed parameter space in
$(\theta_{23}, \Delta m^2)$ space is shown for the same input parameters
as before. Again, both the maximal solution and the wrong octant solution
for $\theta_{23}$ are disallowed at 95\% CL level. While maximality is
still disallowed, a small island opens up around the wrong-octant
solution at the 99\% CL, again to the right of it. Also, $\Delta m^2$
is constrained to a precision of better than 6\% at 3$\sigma$.

\begin{figure}[htp]
\includegraphics[width = \textwidth]{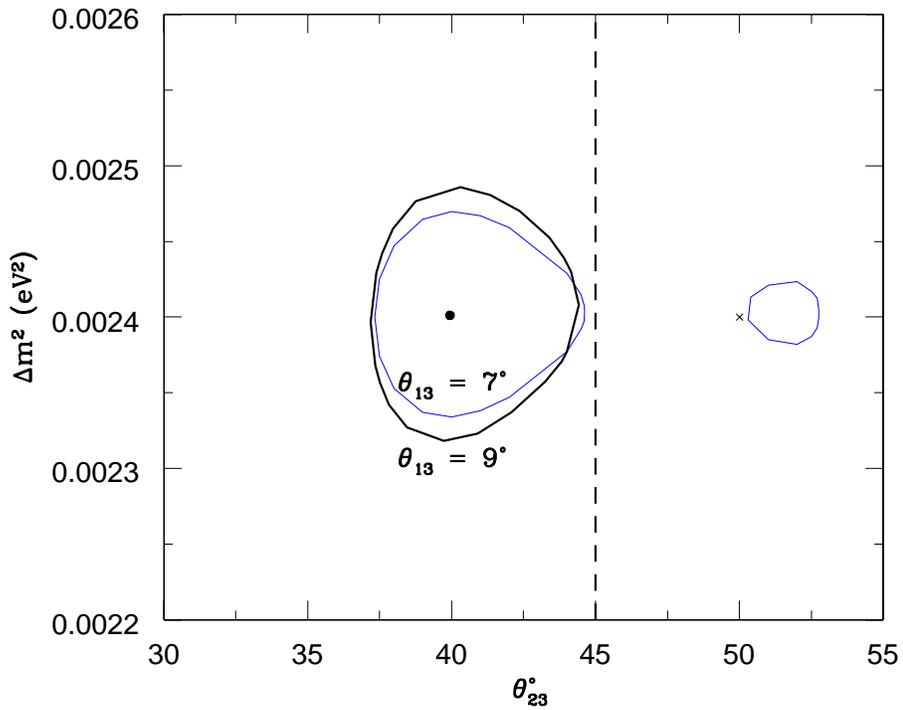}
\caption{As in Fig.~\ref{fig:mdelta37}, but for different $\theta_{13}$
values. Shown are the 99\% CL contours for $\theta_{13} = 7^\circ,
9^\circ$.}
\label{fig:mdelta39}
\end{figure}

The results are a complicated function of both $\theta_{13}$ and $\Delta
m^2$. In general, the wrong-octant and maximal solution is disallowed at
95\% CL up to an input value of $\theta_{23} = 42^\circ$ or $48^\circ$,
i.e., a deviation of $3^\circ$ away from maximal in either direction
(or 10\% deviation in $\sin^2\theta_{23}$) but is allowed at 99\%
CL. This result is mildly dependent on the value of $\theta_{13}$
with results obviously improving for larger $\theta_{13}$. For
instance, the 99\% CL contours for the same input data set are
plotted in Fig.~\ref{fig:mdelta39} for $\theta_{13} = 7^\circ,
9^\circ$. The precision in $\Delta m^2$ is marginally worse for the
larger $\theta_{13}$; however, although at both values of $\theta_{13}$
maximality is disallowed at 99\% CL, the wrong-octant solution is
completely disallowed only for the larger (13) mixing angle.

It is possible to reduce the allowed parameter space by adding the
contribution of the anti-neutrinos as well. That is, we now consider the
input ``data set'' to consist of both the neutrino and anti-neutrino
events ratios. Since we always consider up/down ratios to minimise
errors due to overall normalisation of fluxes and cross-sections,
this still means that we need to separate the charged muons with
good efficiency (that is, identify the process as originating from a
neutrino or anti-neutrino).  In principle, charge misidentification can
lead to systematic errors since, for example, a neutrino event wrongly
identified not only is lost from its parent sample, but also adds to
the anti-neutrino events.  However, in proposed experiments such as
{\sc ical/ino}, charge identification efficiency in this energy range
is better than 98\% so we ignore such correlation errors.

\begin{figure}[htp]
\includegraphics[width = \textwidth]{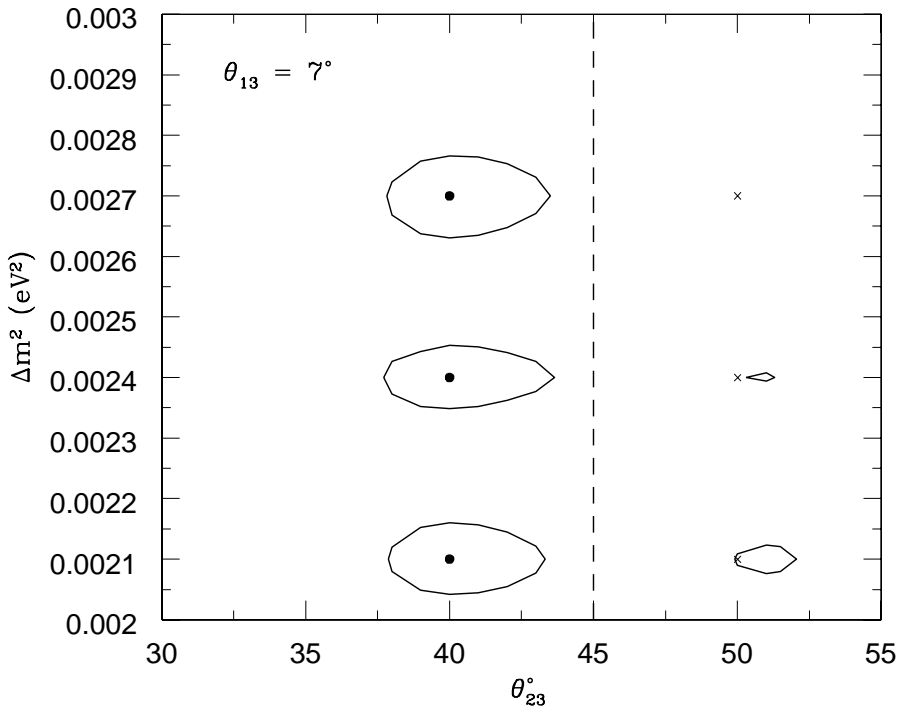}
\caption{As in Fig.~\ref{fig:mdelta37}, for the sum of neutrino and
anti-neutrino up/down event rates, for $(\theta_{23}, \theta_{13})
= (40^\circ, 7^\circ)$ but for a 1$\sigma$ variation in $\Delta m^2$ 
of $\Delta m^2 = (2.1, 2.4, 2.7) \times 10^{-3}$ eV$^2$. While the
wrong-octant solution, that is, the region around $\theta_{23} =
50^\circ$ is disallowed at 95\% CL in all cases (not shown in the
figure), it is allowed at the 99\% CL level as shown. Deviation from
maximality can always be established at 99\% CL.}
\label{fig:mdelta7}
\end{figure}

We show the allowed parameter space on including both the neutrino and
anti-neutrino data sample in Fig.~\ref{fig:mdelta7}. Here the complex
dependence on the input value of $\Delta m^2$ is also shown: as $\Delta
m^2$ is increased between its current 1$\sigma$ allowed values of $\Delta
m^2 \sim (2.1$--$2.7) \times 10^{-3}$ eV$^2$, the allowed parameter space
in $(\theta_{23}, \Delta m^2)$ around the wrong-octant solution shrinks,
and disappears at the upper value of $\Delta m^2$. Hence, it appears
that deviations from maximality as well as determination of the octant
can be typically easily established at the 95\% CL level, provided
$\theta_{13}$ is well-known (and at least of the order $\theta_{13}
= 7^\circ$). Results at the 99\% CL level are harder to quantify and
reflect the complex nature of this many-parameter problem.

We now ask how sensitive these results are to the finite resolution
(in determining neutrino energy and direction, and hence its $L/E$)
of the detector.

\subsection{Event rates with finite detector resolution}

Inclusion of finite detector resolution reduces the sensitivity,
especially beyond the first oscillation minimum and maximum in $L/E$.
The smearing in direction affects the resolution in path-length. A
straightforward way to include such effects is to smear the observed
energy and direction of the neutrino with Gaussian resolution functions:
\begin{eqnarray}
R_1(E', E) & \equiv & \by{1}{\sqrt{{2}{\pi}}\sigma_E} \exp \left[
\by{-(E-E')^2}{2 \sigma_E^2} \right]~; \\ \nonumber
R_2(L', L) & \equiv & \by{1}{\sqrt{{2}{\pi}}\sigma_L} \exp \left[
\by{-(L-L')^2}{2 \sigma_L^2} \right]~.
\label{resfns}
\end{eqnarray}
Hence, the event rate now includes the probability that a neutrino of a
given $L'$ and $E'$ is detected in the detector with path-length
$L$ and energy $E$. We have
\begin{equation}
N^{\alpha,R}_{\rm bin}(x) = \int_{\rm bin}  \d x \int_{E_{\rm min}}
\d E \int \by{\d E'}{E'} R_1(E', E) \int \d L' R_2 (L', L)
{\cal{J}} \frac{\d^2 N^{\alpha}}{\d\ln E'\d\cos\theta' }~,
\label{eventR}
\end{equation} 
where ${\cal{J}}$ is the Jacobean of transformation. 

We re-evaluate the event rates and the up/down ratios using this
equation, with $\sigma_E = 0.15 E'$ and $\sigma_L = 0.15 L'$. These
are realistic widths obtained by a {\sc geant} analysis of atmospheric
neutrino events by both the {\sc monolith} and the {\sc ical/ino}
collaborations. Such a smearing has the effect of accounting for errors in
correctly identifying the $L/E$ bin for a given event, and so accounting
for bin-to-bin correlations.

\begin{figure}[htp]
\includegraphics[width = \textwidth]{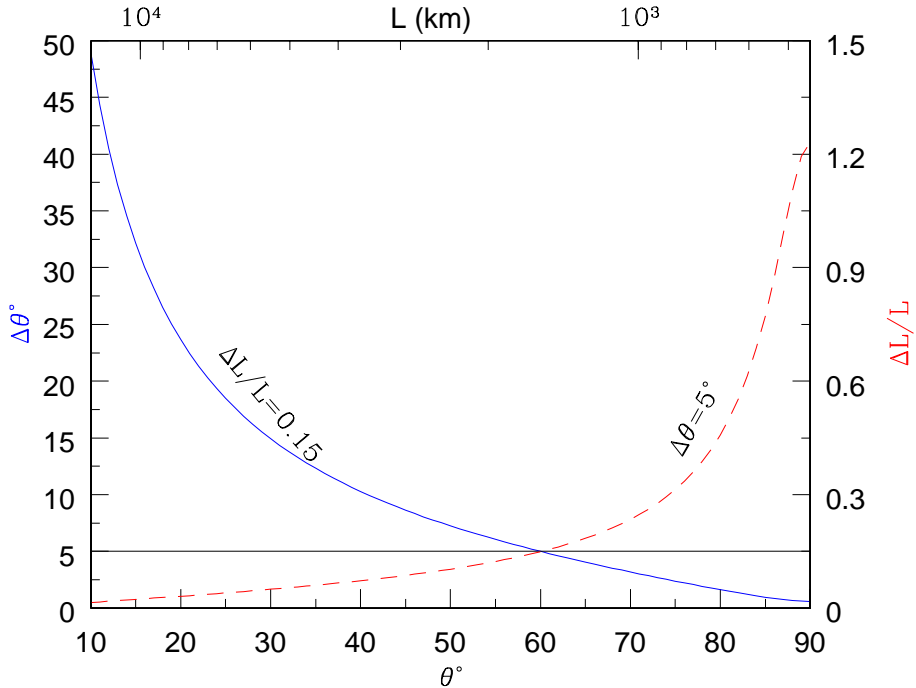}
\caption{The figure shows the effective width in nadir angle of a Gaussian
smearing in the base-length, with width $\Delta L/L = 0.15$, as the
solid line, with axis labelled in $\Delta \theta^\circ$ on the left.
It is seen that a constant relative width in base-length corresponds
to very small angular smearing in the horizontal direction and large
smearing in the vertical direction (small $\theta$ or large $L$). The
dashed line conversely shows the smearing in base-length for a constant
Gaussian width of $\Delta \theta = 5^\circ$, with the scale in $\Delta
L/L$ shown on the axis on the right. It is seen that constant angle and
constant relative widths in base-length are complementary to each other
in their effects. The $x$-axis is labelled both in nadir angle $\theta
(^\circ)$ and base-length $L$ (km) for convenience.}
\label{fig:dtheta}
\end{figure}

We comment in passing on the use of finite widths in $L$ rather
than in $\theta$. The smearing in $L$ actually arises because of the
uncertainties involved in reconstructing the neutrino direction. Hence,
the smearing should be in $\theta$ not $L$. Choosing a resolution in
$\theta$ rather than in $L$ (or equivalently $\cos\theta$) may give
more realistic spreads, especially at large angles where smearing in
$\cos\theta$ may underestimate the large variations in $L$ for small
changes in $\theta$. This can be seen in Fig.~\ref{fig:dtheta}. Applying
a constant $\Delta L/L$ of 15\% corresponds to large angular spreads
at small angles where base-lengths are large, while a constant angular
width $\Delta \theta$ of $5^\circ$ leads to larger spreads in baselines
at large angles. In this paper, we go with convention \cite{monolith,ino}
and define widths in the base-length $L$.

\begin{figure}[htp]
\includegraphics[width = \textwidth]{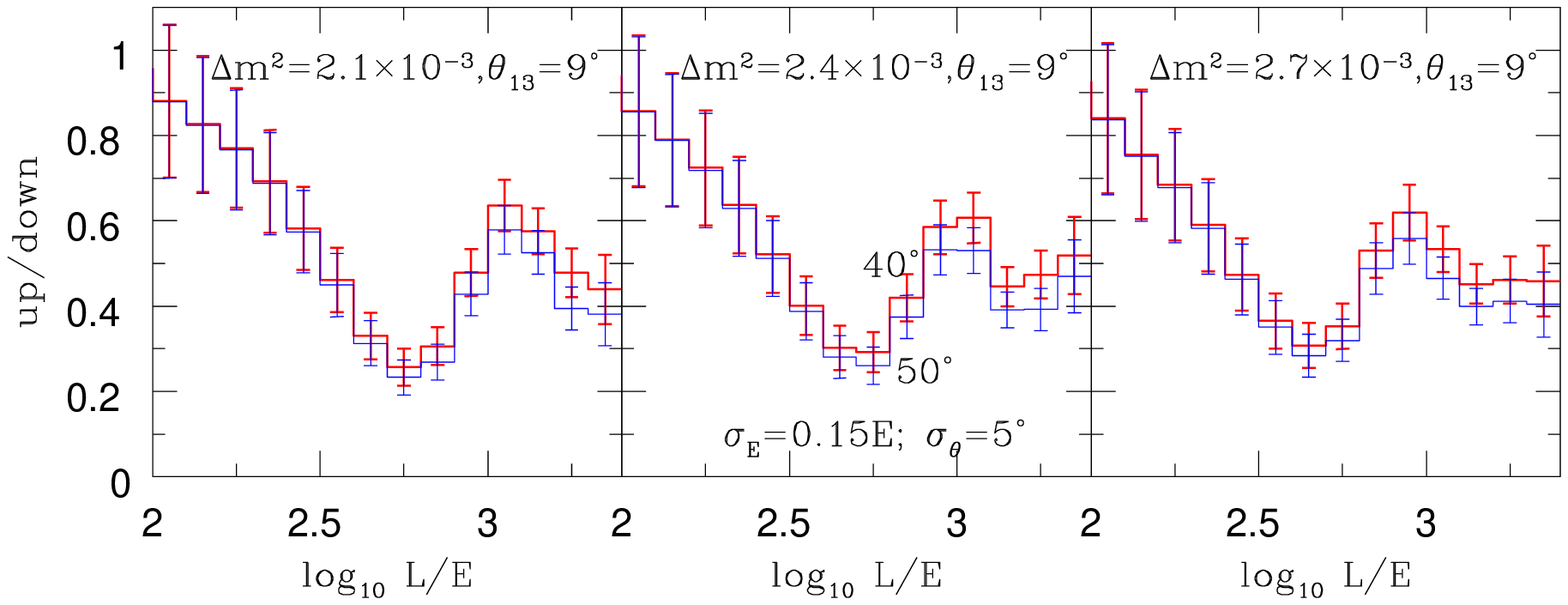}
\includegraphics[width = \textwidth]{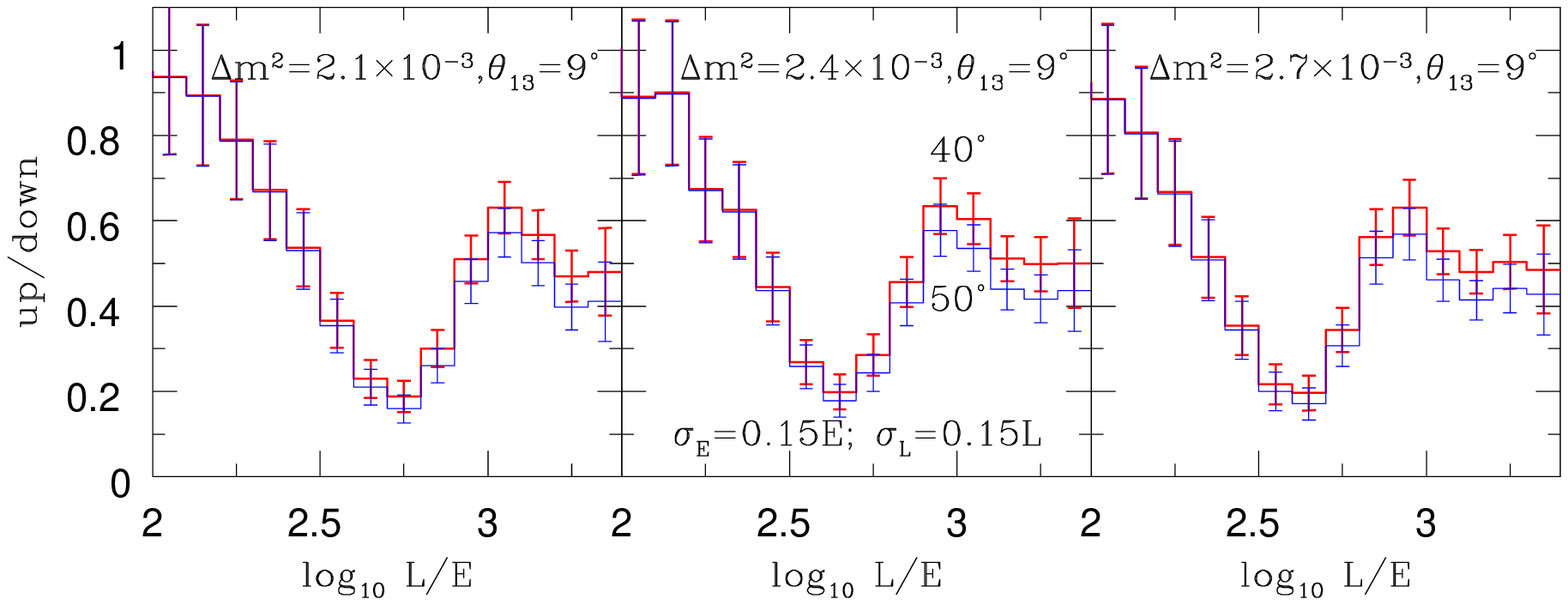}
\caption{As in Fig.~\ref{fig:deltam223}, for $\theta_{13} = 9^\circ$,
including finite detector resolutions. Above: smearing in $E$ and
$\theta$ by Gaussian resolutions with widths $\sigma_E/E = 0.15$ and
$\sigma_\theta = 5^\circ$, and Below: smearing in $E$ and $L$ by
Gaussian resolutions with widths $\sigma_E/E = \sigma_L/L = 0.15$.}
\label{fig:deltam223R}
\end{figure}

In Fig.~\ref{fig:deltam223R} we show the variation of the up/down events
ratio for different $\Delta m^2$ values as a function of $\log_{10}L/E$
for $\theta_{23} = 40^\circ, 50^\circ$ when the resolution function is
included. The upper panels correspond to smearing in $E$ and $\theta$
while the lower panels correspond to smearing in $E$ and $L$. Here
$\theta_{13}$ is fixed to $9^\circ$. A comparison with the similar
Fig.~\ref{fig:deltam223}, where no resolution functions have been
included, shows immediately that the effect of changing the octant
of $\theta_{23}$ remains roughly the same as before (about 1$\sigma$
maximum deviation in each bin) while the oscillation pattern itself gets
smeared away due to finite resolution functions so that the minima and
maxima of the oscillations are not as clearly visible, especially beyond
the first minimum. It is seen that the peaks and troughs are better
defined when smearing in $L$ rather than $\theta$ is used. However, the
differences between smearing in $L$ and $\theta$ otherwise seem marginal.

\begin{figure}[htp]
\includegraphics[width = \textwidth]{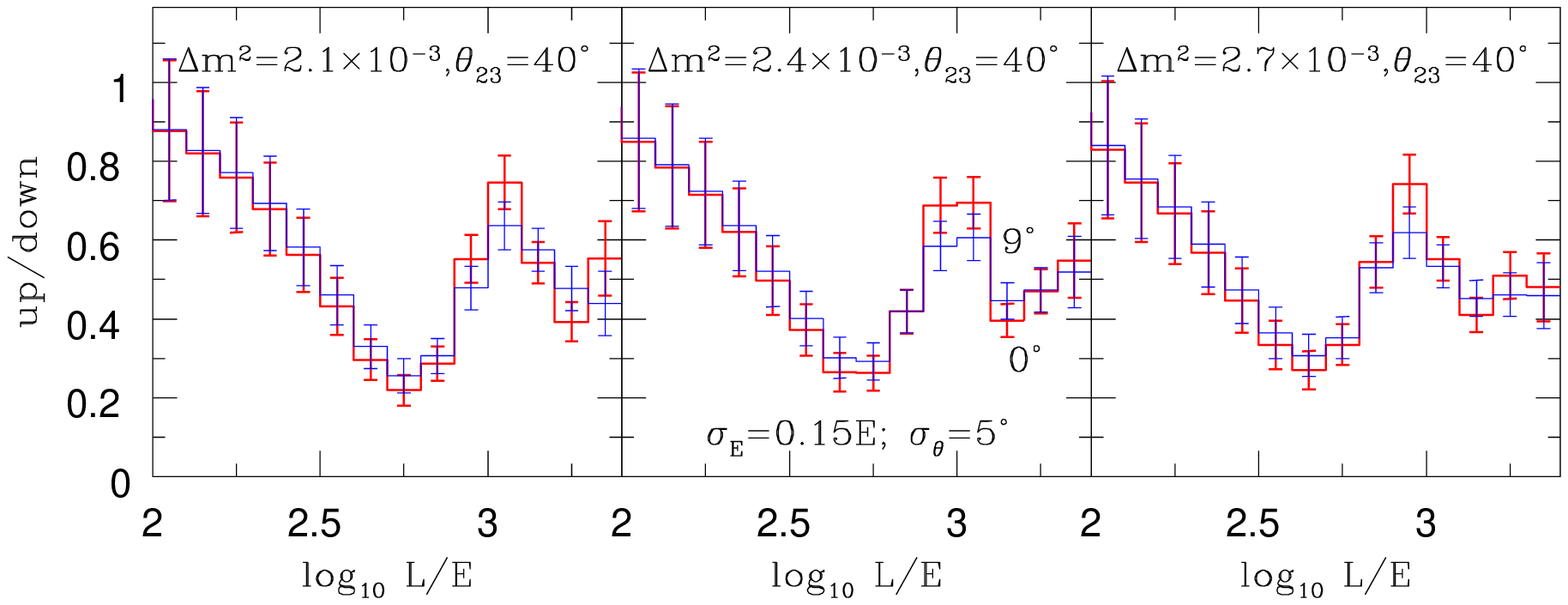}
\includegraphics[width = \textwidth]{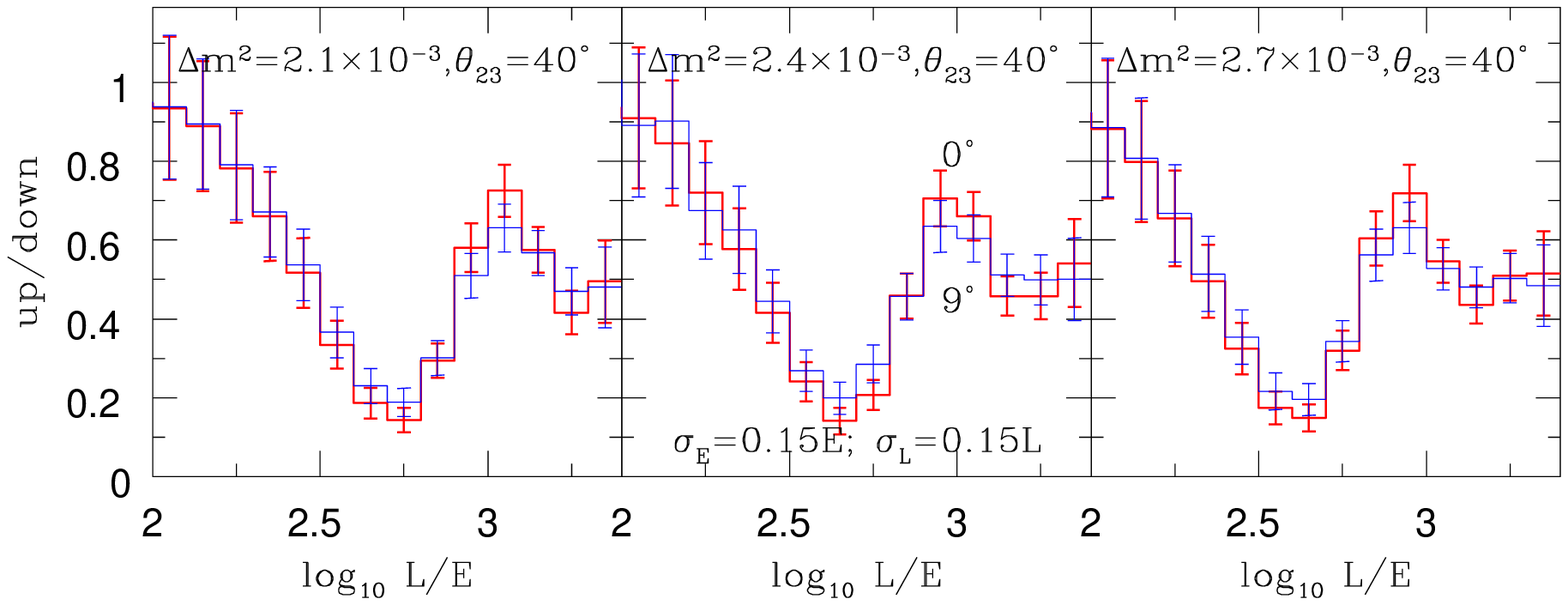}
\caption{As in Fig.~\ref{fig:deltam213}, showing variations with
$\theta_{13} = 0, 9^\circ$, where event rates are now calculated
including finite detector resolution. For more details see the caption
to Fig.~\ref{fig:deltam223R}.}
\label{fig:deltam213R}
\end{figure}

Hence the effect of detector resolution is akin to that of
increasing $\theta_{13}$, which moves both the minima and maxima
of the peaks towards the average value of 1/2. This can be seen in
Fig.~\ref{fig:deltam213R} which shows the variation of the up/down
events, with resolution, at $\Delta m^2 = 2.4 \times 10^{-3}$ eV$^2$,
for different values of $\theta_{13}$. It is clear, especially when
compared with the corresponding results {\em without} resolution function
in Fig.~\ref{fig:deltam213}, that the distinction between the curves for
different $\theta_{13}$ is much reduced. In particular, it is likely
that other experiments, for example, Double-{\sc chooz}, will be able
to resolve $\theta_{13}$ with better accuracy.

\begin{figure}[htp]
\includegraphics[width = \textwidth]{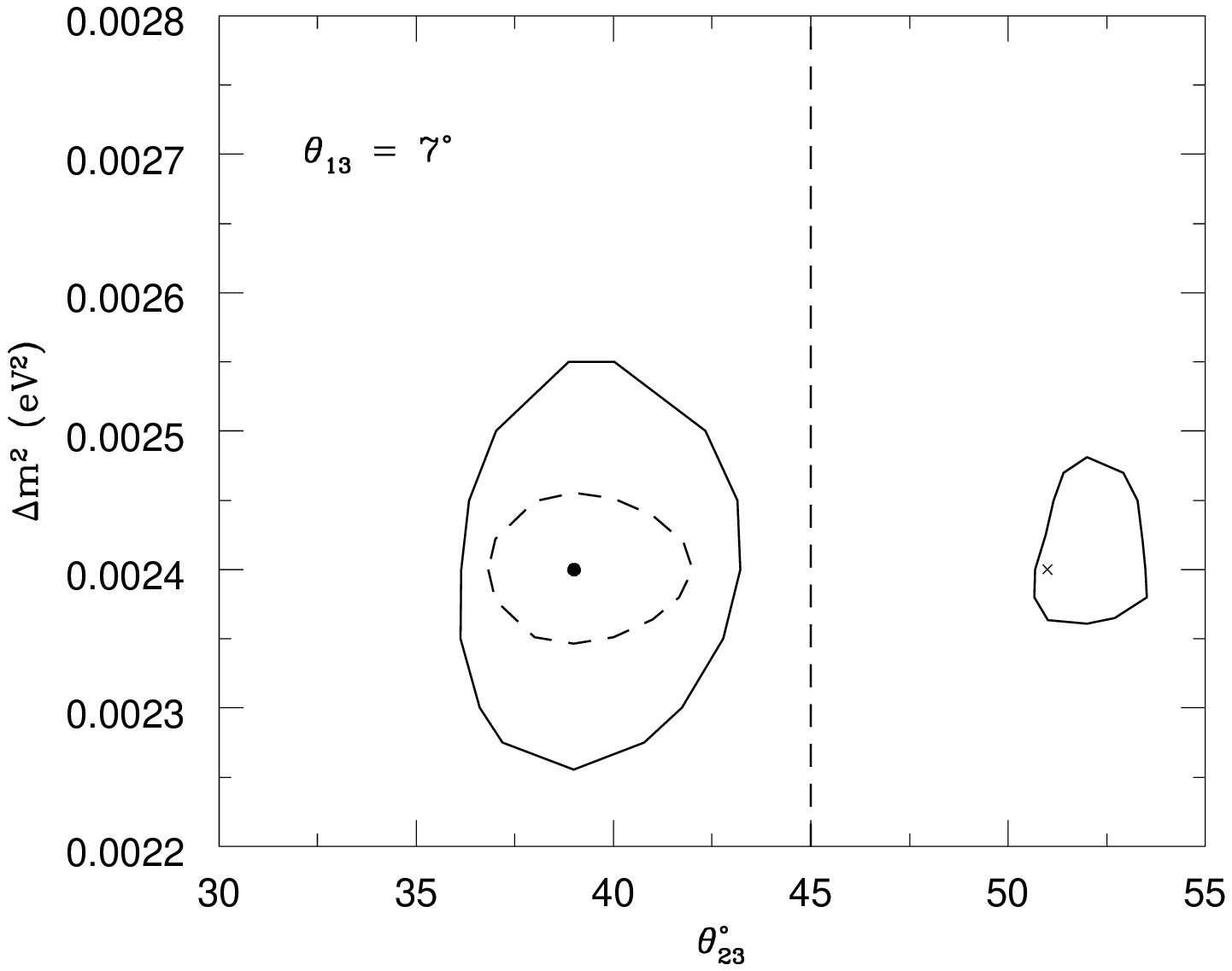}
\caption{Allowed parameter space from neutrino and anti-neutrino
up/down event rates for an exposure of 1000 kton-years and input
parameter values $(\theta_{23}, \theta_{13}) = (39^\circ, 7^\circ)$ for
fixed $\Delta m^2 = 2.4 \times 10^{-3}$ eV$^2$. The 99\% CL contours are
shown for an ideal detector (dashed lines) and for a detector with
finite Gaussian resolutions of widths 15\% in $E$ and $L$. An island of
allowed parameter space near the ``wrong octant'' solution,
$\theta_{23} = \pi/2 - \theta_{23}^{\rm input} = 51^\circ$, marked with a
a cross, is seen.}
\label{fig:mdelta7_39}
\end{figure}

\begin{figure}[htp]
\includegraphics[width = \textwidth]{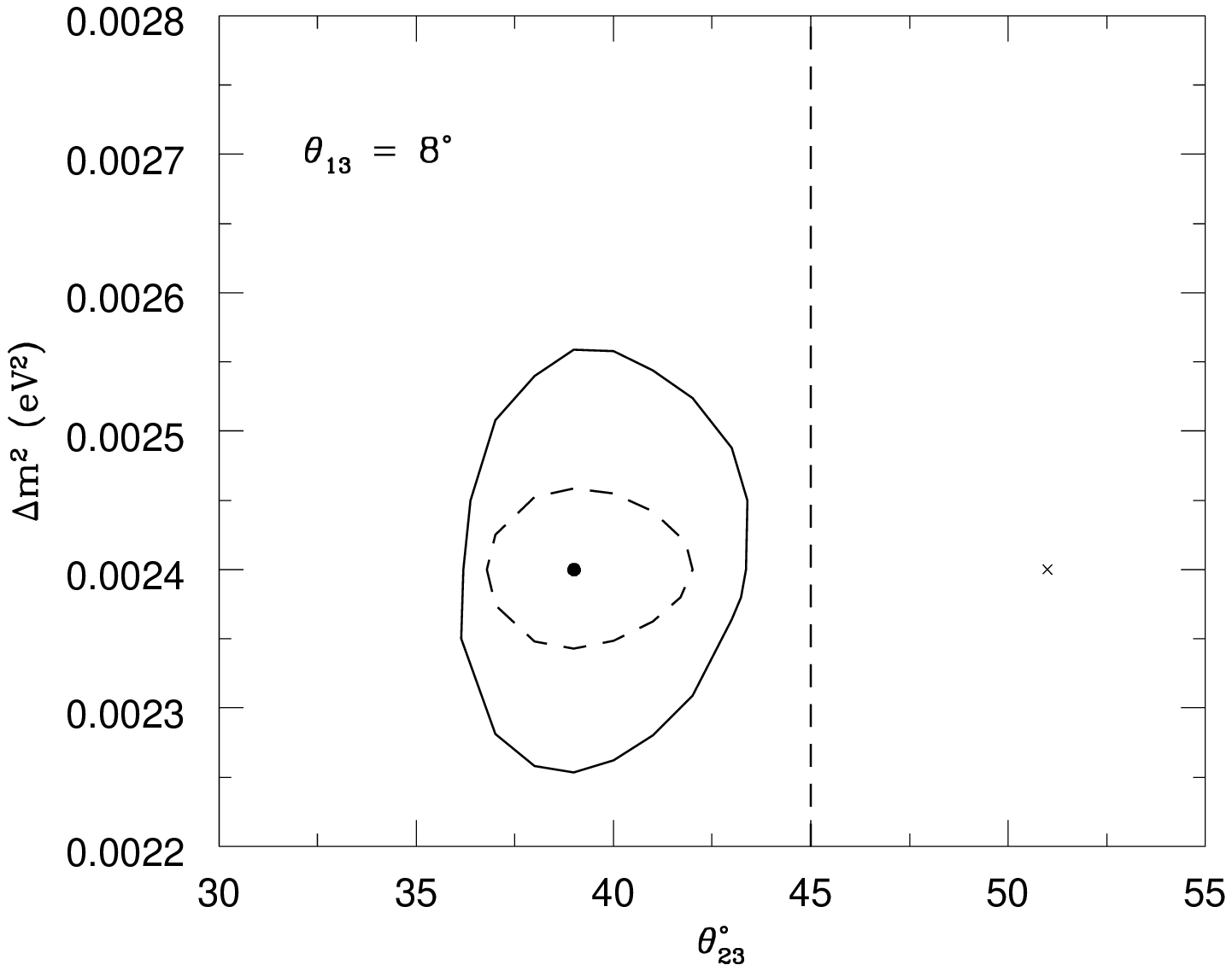}
\caption{As in Fig.~\ref{fig:mdelta7_39} for $\theta_{13} = 8^\circ$.
There is no allowed parameter space near the ``wrong octant'' solution,
$\theta_{23} = \pi/2 - \theta_{23}^{\rm input} = 51^\circ$, marked with a
cross in the figure.}
\label{fig:mdelta8_39}
\end{figure}

We now go on to study the effect of finite detector resolutions on
the extraction of oscillation parameters. We repeat the earlier
calculation, where contours of allowed parameter space were generated
for an input data of up/down neutrino and anti-neutrino event rates in
fixed $L/E$ bins for a set of input values for $(\theta_{13}, \Delta m^2,
\theta_{23})$ and focus, as before, on the octant resolution. We find that
for resolution widths $\sigma_E/E = \sigma_L/L =$ 15\% in both $E$ and
$L$, atmospheric neutrino (and anti-neutrino) data accumulated over
1000 kton-years is sufficient to distinguish a non-maximal solution
from a maximal solution as well as the octant of $\theta_{23}$ for
$\theta_{13} \ge 8^\circ$ and $\theta_{23} \le 39^\circ$ or $\theta_{23}
\ge 51^\circ$. However, the inclusion of finite detector resolution
severely degrades the errors on the parameters. For instance, the error
on $\Delta m^2$ is twice as large as that with an ideal detector. This is
shown for two values of $\theta_{13}$, $\theta_{13} = 7^\circ, 8^\circ$,
for $\theta_{23} = 39^\circ$ ($\sin^2\theta_{23} = 0.4$) and $\Delta
m^2 = 2.4 \times 10^{-3}$ eV$^2$ in Figs.  ~\ref{fig:mdelta7_39} and
\ref{fig:mdelta8_39} respectively.

While maximality in $\theta_{23}$ can be excluded and the octant
distinguished at 95\% CL, only deviation from maximality can be
established at 99\% CL for the smaller $\theta_{13} = 7^\circ$ due to the
island around the ``wrong octant'' solution. It may be mentioned here
that a similar analysis has been performed in Ref.~\cite{probir}. Here
a slightly lower neutrino energy was allowed, down to $E=1$ GeV, although
only DIS interactions were considered. Also, the relevant results in
this paper focus on discriminating the ``right octant'' solution from
the ``wrong octant'' one. However, one must exercise caution in such an
analysis. It is possible, as is very clearly visible in several of the
figures showing allowed parameter space with ``wrong octant'' islands,
that while the exact ``wrong octant'' solution may be disallowed, a
small neighbourhood of this point may still remain allowed. In fact,
it is typically seen that the allowed island at 99\% CL around the
``wrong octant'' solution is typically to the right of the central value.

\begin{figure}[htp]
\includegraphics[width = \textwidth]{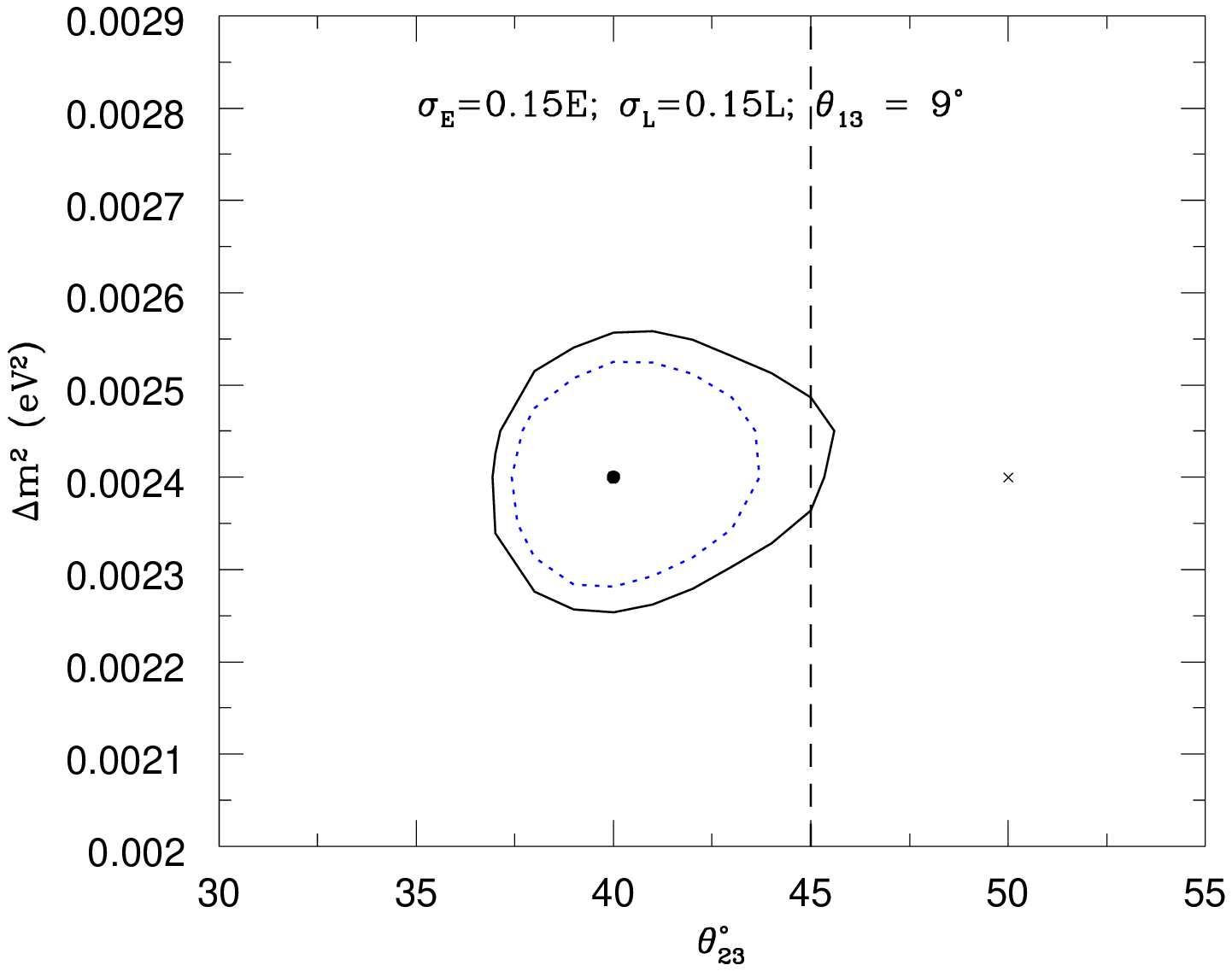}
\caption{Allowed parameter space in $(\theta_{23}, \Delta m^2)$
variables. Input used was $(40^\circ, 2.4\times 10^{-3}$ eV$^2$), for
$\theta_{13} = 9^\circ$, where the up/down ratio was analysed using
finite resolution functions in $E$ and $L$ of widths 15\% each. Shown are
the allowed 95\% and 99\% CL contours. While octant discrimination is
possible at 99\% CL, the issue of maximality can only
be settled at somewhat better than 95\% CL.}
\label{fig:mdeltaR9}
\end{figure}

From Figs.~\ref{fig:mdelta7_39} and \ref{fig:mdelta8_39} it is
seen that a larger $\theta_{13}$ is instrumental in suppressing the
``wrong octant'' solutions but does not shrink the allowed parameter
space around the correct $\theta_{23}$ value. In fact, the allowed
parameter space around $\theta_{23} = 39^\circ$ is virtually the same
in both figures. With larger $\theta_{13} = 9^\circ$, it is possible
to obtain octant discrimination for larger $\theta_{23} = 40^\circ$,
but the issue of maximality can only be settled at little less than 99\%
CL in this case, as can be seen from Fig.~\ref{fig:mdeltaR9}. (Of
course, if deviation from maximality cannot be established, the octant
determination appears to have no meaning. What we mean here is that the
allowed parameter space lies mostly in the first octant. The octant
mirror of the best-fit point is ruled out, but not the maximal value.
We will refer to such solutions as discriminating the octant but not
deviations from maximality.)  It turns out that both octant and
maximality discrimination can be established at 95\% CL for values of
$\sin^2\theta_{23}$ 15\% away from the maximal value. Contrast this with
the result (10\%) of the previous section when an ideal detector is used.

It is clear that resolution functions worsen the precision with which the
magnitude of $\Delta m^2$ and $\theta_{23}$ can be determined but do not
substantially alter the sensitivity to the octant of $\theta_{23}$. That
is, $\theta_{13} = 7^\circ$ ($\sin^2\theta_{13} = 0.015$) remains the
limit at which matter effects are substantial enough to determine the
octant of the (23) mixing angle (as also the mass ordering of the (23)
mass eigenstates, as discussed in Ref.~\cite{IM}), even on including
finite resolution functions, while the actual value of the (23)
angle which simultaneously allows maximality and octant discrimination
at 99\% CL level moves marginally from $\theta_{23} = 40^\circ$ in
the absence of finite resolution effects, to $\theta_{23} = 39^\circ$
($\sin^2 \theta_{23} \approx 0.4$) at the lower limit of $\theta_{13}$.

We note that the situation is not as clean when the true value
of $\theta_{23}$ lies in the second octant.  The muon survival
probability in this case is not as well separated from that for maximal
$\theta_{23}$, unlike when $\theta_{23}$ is in the first octant (see
Fig.~\ref{fig:23t}). Hence, while results for octant discrimination will
be symmetric with the case when $\theta_{23}$ is in the first octant,
the issue of maximality may not be so easily settled. For example,
with other parameters as before, and using $\theta_{23} = 51^\circ$,
which is the octant mirror of $\theta_{23} = 39^\circ$, the ``data''
discriminate between the  right and wrong octant at 99\% CL, but cannot
discriminate the true value of $\theta_{23}$ from maximality. By going
farther away from maximality, for example, at $\theta_{23} = 52.5^\circ$
(and $(\Delta m^2, \theta_{13})$ values of $(2.4 \times 10^{-3}$ eV$^2,
7^\circ$), as before), we once again have maximality discrimination.

\subsection{A note on the inverted hierarchy}

The oscillation probabilities and hence up/down events ratio in
anti-neutrinos with inverted hierarchy are very similar to those
in neutrinos with the normal hierarchy. However, the errors are
about two and a half times as large due to the smaller anti-neutrino
cross-sections leading to correspondingly smaller event rates. Because of
this, a 1000 kton-year exposure is inadequate to address the question of
maximality or octant determination of $\theta_{23}$. Both can be
determined at 99\% CL for an ideal detector for $(\theta_{23},\theta_{13})
= (37^\circ, 9^\circ)$. When resolution functions are included,
only maximality can be established; it is not possible to determine
the octant of $\theta_{23}$ for any value of $\theta_{23}$ allowed by
the Super-Kamiokande results (see Table~\ref{tab:bestfit}). In this, our
results differ from those of Ref.~\cite{probir}. By itself, this result
is not surprising; otherwise it will imply that 500 kton-year exposure
should be adequate for such determinations with normal hierarchy, which
is not the case.

\section{Discussion and Summary}

Neutrino oscillation studies are now moving to the precision determination
of oscillation parameters. The most interesting questions in this
field today centre around the issue of whether the (13) mixing
angle is different from zero and its implications. The impact of this 
fundamental parameter and others were analysed in the first part of this 
paper. In particular we find:

\begin{itemize}

\item $\theta_{13}$ sensitivity: With $\theta_{13}$ nonzero, for
energies beyond 3 GeV, there is a sharp discontinuity in the survival as
well as conversion probabilities, $P_{\mu \mu}$ and $P_{e\mu}$, at the
mantle-core boundary at nadir angle $33^\circ$. (See Fig.~\ref{fig:13t}
and \ref{fig:e13t}). For small path lengths (within the mantle) increasing
$\theta_{13}$ simply decreases (increases) the amplitude of $P_{\mu\mu}$
at the maxima (minima) without significantly altering the oscillation
period.

However, as the paths cross the mantle-core boundary and propagate
within the core, the matter effects significantly modify the period of
oscillations as well, for both the survival and conversion
probabilities. In particular, the jump in $P_{e\mu}$ at the mantle-core
boundary becomes larger and more visible with increasing $\theta_{13}$.
(See Fig.~\ref{fig:e13t}). This important feature can be used with long
baseline neutrinos for Earth tomography studies.

\item $\Delta m^2$ sensitivity: 
A variation in this parameter changes the period of oscillations, even
in the absence of matter effects. For non-zero $\theta_{13}$, with the
matter effects turned on, there are dramatic changes in the probabilities,
as again both the periods and amplitudes are dramatically altered,
especially at around 5 GeV. (See Fig.~\ref{fig:32t}). At around 2 GeV,
a substantial $P_{e \mu}$ at small nadir angles indicates a rather small
$\Delta m^2$, $\Delta m^2 \lesssim 2 \times 10^{-3}$ eV$^2$. (See
Fig.~\ref{fig:e32t}).

\item $\theta_{23}$ sensitivity: For $\theta_{13}$ nonzero, we note that
as $\theta_{23}$ increases from the first octant through to the second
octant, the survival probability systematically decreases for most nadir
angles in the energy range 5--10 GeV. (See Fig.~\ref{fig:23t}). However,
the conversion probability, $P_{e \mu}$ systematically increases as
$\theta_{23}$ increases at all nadir angles and all energies. This is not
very significant for atmospheric neutrinos, but can be studied separately
with long baseline neutrinos.

\end{itemize}

Finally, the probabilities are sensitive to variations in the {\sc prem}
profile (within allowed limits) as also the CP phase. However, these
dependences are rather small and insignificant and are unlikely to be
measured with atmospheric neutrino studies of the type discussed here.
The results discussed for CP phase in this paper can therefore be seen
only with long baseline neutrinos when all other parameters are
presumably known to good precision.

The probabilities $P_{\mu\mu},P_{e\mu}$ are then used to analyse
the atmospheric muon neutrino events to determine the deviation from
maximality and/or the octant of the (23) mixing angle $\theta_{23}$.
Such dependence arises purely from matter effects as the neutrinos
propagate through Earth. Since matter effects are different in the
case of neutrinos and anti-neutrinos, we assume a detector with charge
discrimination capability, such as the proposed {\sc ical/ino} detector,
to heighten the sensitivity to such matter-dependent effects. Moreover,
all matter terms are proportional to $\sin^2\theta_{13}$ so we assume
that this (13) mixing angle is different from zero.

We study the up/down events ratios, where by up (down)-neutrinos, we
mean neutrinos arriving at the detector from below (above), so their
nadir angles range from 0--$90^\circ$ (90--$180^\circ$). Analysis of
ratios of events rather than events themselves significantly reduces
errors from overall normalisation of fluxes and cross-sections; the
former can be as large as 30\% and the latter around 10\%. Smearing in
energy $E$ and base-length $L$ by use of Gaussian functions is used to
simulate finite detector resolutions. Typical relative widths of 15\%
are used for both energy and base-length. The allowed parameter space
obtained on fitting up/down neutrino and anti-neutrino events ratios in
any two of $(\Delta m^2, \theta_{23}, \theta_{13})$ is considered.

Since the up/down events ratio has very different dependence on
$\theta_{23}$ and $\theta_{13}$, in principle, they can be simultaneously
determined from such studies. The proposed detectors such as {\sc
ical/ino} with charge identification capability may be in principle
suitable for determination of these parameters. In practice, however,
these detectors are not well suited for precision determination of
$\theta_{13}$ with atmospheric neutrinos. A precise measurement of this
parameter is likely to be made, for instance, by the Double-{\sc chooz}
experiment.

We summarise below our observations on the simultaneous determination
of $\vert \Delta m^2 \vert$ and $\theta_{23}$ from an analysis of the
event rates at these detectors with focus on the determination of
$\theta_{23}$. We distinguish three scenarios. The best case is when
both deviation of $\theta_{23}$ from maximality as well as its octant
can be established.  In some cases, a maximal value of $\theta_{23}$
is disallowed, so deviations from maximality can be established but the
octant discrimination may not be possible. The third is where the {\em
best-fit} value lies in one of the octants and the region in parameter
space around the mirror to the best-fit value in the other octant is
ruled out at the precision under consideration; however, the maximal
value is not ruled out at this precision. In such a case, the allowed
parameter space still lies mostly in one octant and hence we consider
that we have octant but not maximality discrimination.

\begin{itemize}

\item If $\theta_{13}$ is known precisely in the near future, as is
likely from Double-{\sc chooz}, then in the energy range $E = 5$--10 GeV,
where matter effects are largest, the data from {\sc ical/ino} will be
able to study deviations of $\theta_{23}$ from maximality as well
as determine the octant of this angle, provided $\theta_{13} \ge
7^\circ$.

\item It must be emphasised that the octant discrimination is more
easily done than establishing deviation from maximality for larger
$\theta_{13}$ and the reverse is true for smaller $\theta_{13}$ when
islands of allowed parameter space begin to appear near the ``wrong
octant'' solution. Also, values of $\theta_{23}$ in the first octant are
more easily distinguished from maximality than those in the second
octant. This result is in contrast to those obtained in
Ref.~\cite{probir}.

\item In particular, deviations from maximality and octant discrimination
to 99\% CL can be obtained if $\sin^2\theta_{23} \le 0.4$
and $\sin^2\theta_{13} \ge 0.015$. These studies used standard Gaussian
resolution functions with widths $\delta L/L = \delta E/ E =$ 15\%
to smear the events. Similarly, we must have $\sin^2\theta_{23} \ge
0.63$ for corresponding results in the second octant.

\item In an earlier paper\cite{IM}, it was also shown that the same
processes are also eminently suited to determine the (23) mass ordering.
Both these determinations need large exposures of roughly 1000
kton-years as well as, crucially, the charge discrimination.

\item As discussed in \cite{IM}, the issue of determining the (23) mass
ordering and hence establishing the neutrino mass hierarchy
is best done using the difference asymmetry defined in
Eq.~\ref{eq:da} which is the {\em difference} of the up/down events
ratios with neutrinos and anti neutrinos. For the octant discrimination
that is the crux of this paper, the relevant observable is in fact the
{\em sum} of these two ratios. 

\item The same experiment can therefore study both these questions,
while requiring large exposures, 1000 kton-years, in both cases. Hence,
studies of neutrino oscillations with atmospheric neutrinos, while being
difficult, are probably the only source of precision measurements, at
least until very large megaton detectors and/or neutrino factories
become a reality.

\end{itemize}
Even so, it is important to point out that the determination of
the hierarchy is likely to be relatively easier than the octant
determination. This is because, as seen from Fig.~\ref{fig:lebin}, the
octant effect rides on the hierarchy issue as a sub-dominant effect.

Another way to see this is that the octant determination is dominated by
the (anti) neutrino events ratio in the (inverted) normal hierarchy,
which are the relevant sectors where matter effects dominate.
Since the statistics is smaller by half for anti-neutrinos for the same
exposure, the significance of the results deteriorates in the case of
inverted hierarchy.

The bulk of the results presented in this paper therefore pertain to
the normal mass hierarchy. Results of similar significance can only be
established with the inverted mass hierarchy if the exposure is at least
twice that for normal hierarchy. Hence, such detectors, with exposures
of 1000 kton-year or so, may be able to settle the issue of mass hierarchy
but, if the ordering is inverted, the issue of maximality of $\theta_{23}$
may remain an open question.

\vspace{0.2cm}

\noindent{\bf Acknowledgements}:
We are grateful to Dave Casper for making the {\sc nuance} software freely
available, and answering a long list of questions on its use. The work
of N.S. was partially supported by the Department of Science and
Technology, India.

\section{Appendix}

One of the earliest analytical methods for the calculation of oscillation
probabilities with variable matter density is in Ref.~\cite{rajaji}. The
oscillation probability was derived assuming matter to be made up of a
series of slabs through which neutrinos and anti-neutrinos propagate. Each
slab has smoothly varying density, but the density itself has discrete
jumps at the junction of adjacent slabs. In a recent paper, Akhmedov et
al.~\cite{akhmedov}, have provided an excellent collection of approximate
analytic formulae for the neutrino oscillation probabilities. Exact
analytical formulae have also been derived in the case of vacuum
\cite{vacuum} and for matter with constant density slabs \cite{matter}. Though
complicated, exact analytic formulae have been derived for non-uniform
density with linear \cite{linear} and exponentially \cite{expo} varying
density. Approximate analytic formulae with varying assumptions have
been derived in a number of papers \cite{approx}.

Here, we use a standard Runge-Kutta solver to numerically propagate the
neutrinos through Earth's matter, using the {\sc prem} Earth density
profile for a spherical equivalent Earth with radius $R_E = 6371$ km. We
outline some details about the numerical calculation. We also highlight
how results from an algorithm using constant density slabs differ from
those with the {\sc prem} profile. Our technique is to numerically evolve
the flavour eigenstates using the equation,
$$
i \by{\d } {\d t} \left[ \nu_\alpha \right] = \by{1}{2E}\,\left( U M^2 
U^\dagger + {\cal A} \right) \left[ \nu_\alpha\right]~,
\eqno(6)
$$
where ${\cal A}$ is the diagonal matrix, diag$(A, 0, 0)$ and
$\left[\nu_\alpha\right]$ denotes the vector of eigenstates, $\nu_\alpha$,
$\alpha=e,\mu,\tau$. As usual, $U$ is the MNS mixing matrix defined in
Eq.~\ref{eq:mns}, $M^2$ is the mass-squared matrix with the
diagonal piece proportional to $m_1^2$ removed: $M^2 =$ diag$(0,
\delta_{21}, \delta_{31})$. Note that in a constant density slab approximation,
the mass eigenstates are propagated, including the ``sudden'' jumps
across the density discontinuities.

\begin{figure}[htp] 
\includegraphics[width = \textwidth]{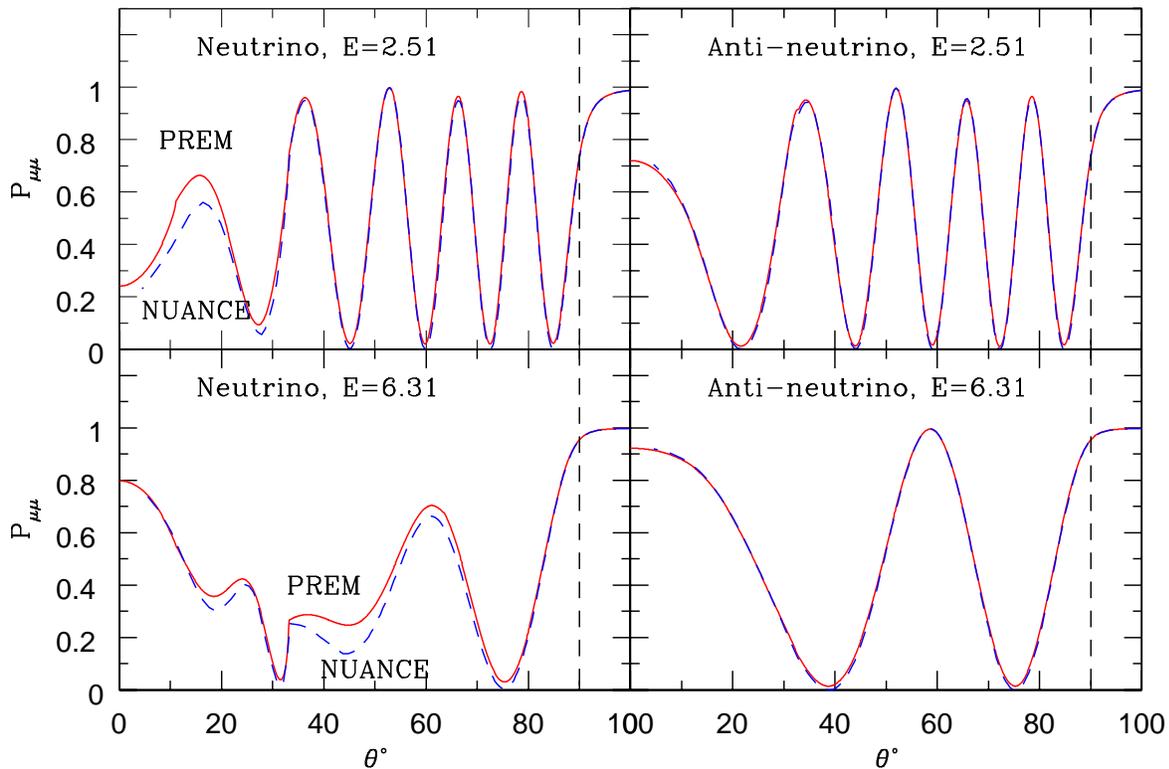}
\caption{The muon neutrino and anti-neutrino survival probability as a
function of nadir angle for different energies. Shown are the results
from an exact numerical method using the {\sc prem} Earth density profile
(solid lines) and a constant density slab model from {\sc nuance}
\cite{nuance}.}
\label{fig:profile} 
\end{figure}

As mentioned in the text, the chief difference between the constant
density approximation and the {\sc prem} profile lies in the resonance
effects. This can be seen from Fig.~\ref{fig:profile} where the results
of using the {\sc prem} profile in our calculations, and the {\sc
nuance} \cite{nuance} result for the same parameters, using constant
density slabs, is shown. Here we have set the parameters $\delta_{21}$,
$\Delta m^2$ (normal hierarchy), $\theta_{21}$ and $\theta_{23}$ to their
best-fit values as given in Table~\ref{tab:bestfit} and set the CP phase
to zero. We use $\theta_{13} = 9^\circ$. For a low value of
$E = 2.51$ GeV, where resonance occurs in the core, the constant density
slab of {\sc nuance} does not show resonance since the precise density
value needed for resonance does not occur. Of course, matter effects are
present and are large; however, this difference causes the two curves for
the neutrino survival probability to begin to separate in this region,
as can be seen from the top left panel of the figure. It is seen that
the anti-neutrino probabilities, which are relatively insensitive to
matter effects in the normal hierarchy, however, match exactly.

At larger energies, $E = 6.31$, the ``missed'' resonance occurs in
the mantle itself. Hence, deviations between the two curves are seen
at almost all zenith angles. The difference can be up to 10--15\%.
Note also that the prominent effect of core-crossing, which occurs at
$33^\circ$, can be seen clearly in both curves.  Again, the anti-neutrino
probabilities match.

We have also checked that, by taking smaller and smaller slabs, and
approximating the {\sc prem} profile as closely as possible, the results
from the constant density slab approach those using the exact Runge-Kutta
solver with {\sc prem} profile. However, the resultant increase in number
of slabs is quite large.

\end{document}